\documentclass[
 aip,
 jmp,
 amsmath,amssymb,
 reprint,
]{revtex4-1}

\usepackage[english]{babel}
\usepackage{graphicx}
\usepackage{amsmath,amssymb,amstext,amsfonts}
\usepackage{xfrac}

\graphicspath{{./}{./figs/}{./plots/}}
\DeclareGraphicsExtensions{.png,.eps,.pdf,.jpg}

\renewcommand{\vec}[1]{\boldsymbol{#1}}

\usepackage{hyperref}
\hypersetup{
    colorlinks=true,
    linkcolor=blue,
    filecolor=blue,      
    urlcolor=blue,
    citecolor=blue,
}
 
\begin{document}

\title{Wakefield-Induced Ionization injection in beam-driven plasma accelerators}

\author{A. Martinez de la Ossa}
\affiliation{Deutsches Elektronen-Synchrotron DESY, D-22607 Hamburg, Germany}
\affiliation{Institut f\"{u}r Experimentalphysik, Universit\"{a}t Hamburg, D-22761 Hamburg, Germany}
\author{T.~J. Mehrling}
\affiliation{Deutsches Elektronen-Synchrotron DESY, D-22607 Hamburg, Germany}
\author{L. Schaper}
\affiliation{Deutsches Elektronen-Synchrotron DESY, D-22607 Hamburg, Germany}
\author{M.~J.~V. Streeter}
\affiliation{Deutsches Elektronen-Synchrotron DESY, D-22607 Hamburg, Germany}
\author{J. Osterhoff}
\affiliation{Deutsches Elektronen-Synchrotron DESY, D-22607 Hamburg, Germany}

\date{\today}

\begin{abstract}
We present a detailed analysis of the features and capabilities of Wakefield-Induced Ionization (WII) injection in the blowout regime of beam driven plasma accelerators.
This mechanism exploits the electric wakefields to ionize electrons 
from a dopant gas and trap them in a well-defined region of the accelerating 
and focusing wake phase, leading to the formation of high-quality witness-bunches~\cite{delaOssa:2013tma}. 
The electron-beam drivers must feature high-peak currents ($I_b^0\gtrsim 8.5~\mathrm{kA}$) 
and a duration comparable to the plasma wavelength to excite plasma waves in the blowout regime and enable WII injection. 
In this regime, the disparity of the magnitude of the electric field in the driver region
and the electric field in the rear of the ion cavity allows for the selective ionization and subsequent trapping 
from a narrow phase interval. 
The witness bunches generated in this manner feature a short duration and small values of the
normalized transverse emittance ($k_p\sigma_z \sim k_p\epsilon_n \sim 0.1$). 
In addition, we show that the amount of injected charge can be adjusted by tuning the concentration 
of the dopant gas species, which allows for controlled beam loading and leads to a reduction of the total energy spread of the witness beams. 
Electron bunches, produced in this way, fulfil the requirements to drive blowout regime plasma wakes at a higher density and to trigger WII injection in a second stage. 
This suggests a promising new concept of self-similar staging of WII injection in steps with increasing plasma density,
giving rise to the potential of producing electron beams with unprecedented energy and brilliance from plasma-wakefield accelerators.  

\end{abstract}

\pacs{52.40.Mj, 52.65.Rr, 52.25.Jm, 52.59.-f}

\maketitle

\section{Introduction}
\label{sec:int}
Beam driven plasma wakefield accelerators (PWFA)~\cite{veksler:1956,Chen:1985pwfa} 
can generate and sustain accelerating gradients in excess of $\sim10~\mathrm{GV/m}$ 
over meter-scale distances. 
This was proven in a landmark experiment at SLAC~\cite{Blumenfeld:2007aa}, 
where the energy of the electrons in the tail of a $42~\mathrm{GeV}$ electron beam
was doubled within a distance of less than a meter, 
after being accelerated in the plasma wake, which was excited by the head of the beam.
Harnessing such extreme fields for the production of multi-$\mathrm{GeV}$ energy,
high-brightness electron beams will enable a new generation of accelerators,
capable of compactly driving applications e.g.~in particle physics, optics, medicine or materials science.

Improved control over the injection of beams into a suitable phase of the plasma wake
is a necessary step towards the production of beams from plasma-based accelerators
with a quality comparable to those generated in current state of the art particle accelerators.
The first peaked electron beam spectra were obtained from laser driven plasma wakefield accelerators (LWFA)~\cite{Tajima:1979bn} 
by uncontrolled nonlinear wavebreaking~\cite{Mangles:2004uq,Geddes:2004fk,Faure:2004fk}.
Subsequently, the implementation~\cite{Chien:2005prl, Faure:2006vn, Rechatin:2009prl, Pak:2010zza, McGuffey:2010prl, Clayton:2010, Gonsalves:2011fk}
of earlier proposed~\cite{Umstadter:1996uq,Esarey:1997cp,bulanov:1998, Chen:2012pop} controlled injection methods
paved the way towards electron beams with improved tunability, stability and quality.
The experimental realization of injection techniques in PWFA is not yet as developed as in LWFA,
and only recently PWFA has been advanced from single-beam electron acceleration~\cite{Blumenfeld:2007aa,Hogan:2005prl}
to the acceleration of distinct witness beams~\cite{Litos:2014}. 
To obtain high-quality beams from PWFA, different injection techniques, based on density tailoring~\cite{Suk:2001ddr},
magnetic-induced trapping~\cite{Vieira:2011qn} or external witness bunch generation~\cite{Litos:2014} have been proposed.
A promising approach is the injection of electrons in plasma by means of 
field-induced ionization of a dopant gas with appropriate ionization potential~\cite{Oz:2007mn, Kirby:2009zzb}.
Injection from ionization, induced in a dopand gas by the radial electric field of the driving beam was first observed in an experiment~\cite{Oz:2007mn} at FACET~\cite{Hogan:FACET}.
That method is highly sensitive to details of the initial charge density distribution 
and the transverse emittance of the driving beam and is therefore not easily controlled and contingent on fluctuations of the beam structure. 
In addition, the injection of excessive charge can severely beam load the wake and may result
in the production of a continuous current that distorts the wakefield~\cite{Vafaei-Najafabadi:2014prl}
instead of a well defined witness beam. 
Laser-triggered ionization injection~\cite{Hidding:2012th,Li:2013} 
offers the possibility to control the volume of injection in a precise way. 
These methods rely on a high degree of spatial ($\sim 1 \, \mu\mathrm{m}$) and temporal ($\sim 10 \, \mathrm{fs}$)
alignment of the laser pulses with respect to the electron beam in order to induce ionization in the desired phase of the plasma wake and to produce high quality beams.

A novel ionization-based injection strategy, Wakefield-Induced Ionization (WII) injection,
was recently proposed~\cite{delaOssa:2013tma} as a method to achieve
a high degree of control over the injection of electrons into the appropriate phase of the plasma wave.
In contrast to the aforementioned methods, WII injection is neither sensitive to details of the beam structure,
nor does it rely on additional devices, such as lasers, for the injection. 
Instead, it exploits the difference of the absolute electric-field strength between the accelerating and decelerating regions of the first wakefield bucket in the blowout regime~\cite{Rosenzweig:1991yx,Lotov:2004zz,Lu.PRL.96.165002} to selectively ionize a small volume of a background dopant gas near the phase of maximum acceleration. 
In this way the production of high-quality, ultra-short ($\sim\mathrm{fs}$), 
low-emittance ($\sim\mathrm{\mu m}$), multi-$\mathrm{GeV}$-energy electron beams 
from a comparably simple experimental setup is possible.

In this work, we present a detailed analysis of the features and capabilities of WII injection.
Since the method relies on properties which are inherent in the structure of beam-driven blowout plasma wakes,
we first review the phenomenology of PWFA in the blowout regime (Sec.~\ref{sec:wiii}).
The most important scalings are discussed there and compared to data from
3D simulations with the Particle-In-Cell (PIC) code OSIRIS~\cite{Fonseca:osiris,Fonseca:2008,Fonseca:2013}.
This analysis allows for a thorough determination of the requirements for WII injection. 
The injection principle is demonstrated in exemplary 3D simulations with OSIRIS, 
considering two realistic experimental scenarios for PWFA, 
namely the FLASHForward project at DESY (Sec.~\ref{sec:FLASHForward}) 
and the FACET experiment at SLAC (Sec.~\ref{sec:FACET}). 
WII injection also allows controlled beam-loading to reduce the total energy spread
of the witness bunch (Sec.~\ref{sec:beamload}). 
This results in the generation of compact, high-energy, low-emittance and high-current witness bunches, 
which fulfil all requirements to again trigger WII injection as driver beams in substantially higher density plasmas.
This brings up a new concept of self-similar staging, which has the potential of producing electron beams 
with unprecedented energy and quality in PWFA (Sec.~\ref{sec:sss}). 

\section{Theory}
\label{sec:wiii}
\begin{figure}[!t]
 \centering
  \includegraphics[width=1.0\columnwidth]{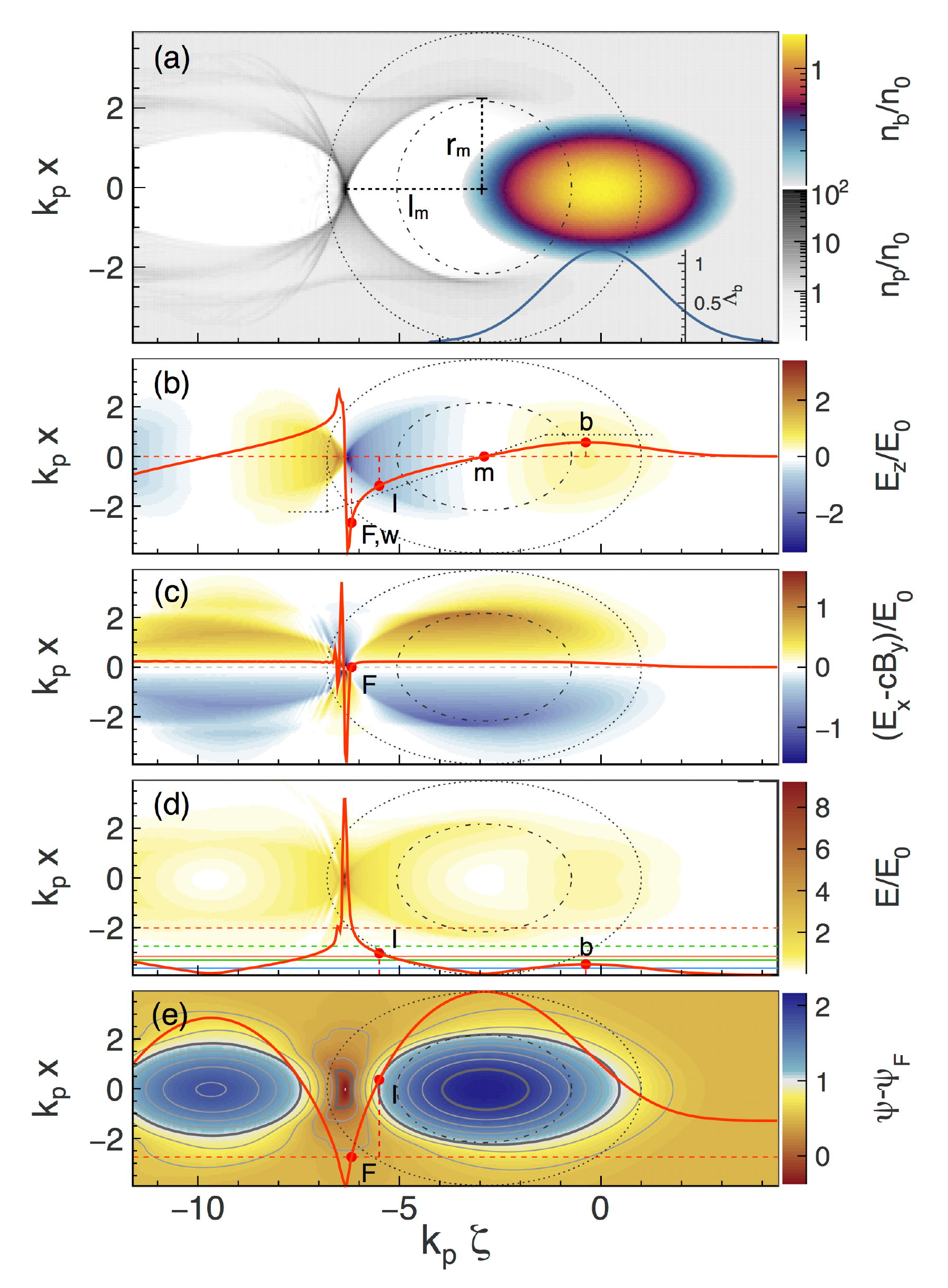}
  \caption{
    An OSIRIS 3D simulation of a high-current ($I_b = 10~\mathrm{kA}$), moderately wide (($k_p\sigma_r = 0.8$)), axially symmetric Gaussian electron beam,
    going through a plasma at the (linear) resonant length ($k_p\sigma_z = \sqrt{2}$).
    (a) Spatial particle density. 
    (b) Longitudinal electric field. (c) Transverse wakefield. (d) Electric field magnitude. (e) Wake potential.
    Red solid lines are the corresponding quantity along the on-axis region, except for (c) and (d),
    where the profile is taken $0.1~k_p^{-1}$ off-axis.
    The outer and inner circles represent the blowout radius estimations through Eq.~(\ref{eq:radiuslo}) and Eq.~(\ref{eq:radiuslu}), respectively.
    The dark dotted lines in (b) indicate the model estimations for the maximum decelerating field in the beam region (right horizontal line, Eq.~(\ref{eq:ezbmax})),
    the maximum accelerating field (left horizontal line, Eq.~(\ref{eq:ezmax})), 
    and the longitudinal field slope around the center of the cavity (diagonal line, Eq.~(\ref{eq:ezcenter})). 
  }
\label{fig:wii10ka} 
\end{figure}
PWFA use relativistic charged particle beams that are sufficiently dense  
to significantly displace the plasma electrons from their ions by means of the Coulomb force of its charge. 
As the beam passes, the displaced plasma electrons are attracted back 
by the excess of positive ions left behind, and oscillate around their equilibrium position,
generating in this way a plasma density perturbation that is propagating at the velocity of the beam ($v_b \approx c$).
For small displacements, these oscillations are harmonic at a frequency given by
$\omega_{p} = \sqrt{n_0 e^2/m\epsilon_{0}}$ and a wavelength of $\lambda_{p} = 2 \pi / k_p = 2 \pi c / \omega_p$,
where $n_0$ is the plasma particle density, $\epsilon_0$ is the vacuum permittivity,
$c$ the speed of light, and $m$ and $e$ are the electron mass and charge, respectively.
High-current ($I_b \gtrsim 1~\mathrm{kA}$), resonant-length ($L_b \approx \lambda_p$) and narrow ($k_p\sigma_r \lesssim 1$) electron drivers  
blow out all the plasma electrons from their propagation path,
creating an ion cavity with no plasma electrons inside. 
As can be seen in PIC simulations (Figure~\ref{fig:wii10ka}~(a)), 
this ion-cavity (or blowout) is clearly delimited by a sheath of plasma electrons, 
which accumulate at a distance $r_l$ from the propagation axis.
The blowout regime features ideal properties for the acceleration and transport of compact electron bunches~\cite{Rosenzweig:1991yx}.
Given the absence of radial currents inside the blowout (and assuming cylindrical symmetry),
it can be directly shown from Maxwell equations that
inside the ion cavity 
the transverse wakefields, $W_\perp \equiv E_r-cB_\phi$, depend linearly on the radius $W_\perp/E_0 = k_pr/2$,
and are constant along the co-moving variable $\partial_\zeta W_\perp = 0$, where the co-moving variable is defined as $\zeta = z - ct$ (see Figure~\ref{fig:wii10ka}~(c));
while the longitudinal wakefield, $W_z \equiv E_z$, is constant along the radius in the blowout, $\partial_r E_z = 0$ (see Figure~\ref{fig:wii10ka}~(b)).
Here $E_0 \equiv k_p mc^2/e$ is the cold non-relativistic wavebreaking field~\cite{Dawson:1959}.
The longitudinal structure of $E_z$, on the other hand, exhibits a substantial difference in magnitude 
between the decelerating part, at the region of the driver, $E_z^b = E_z(\zeta_b)$, and the accelerating part,
at the rear of the blowout, $E_z^w = E_z(\zeta_w)$ (Figure~\ref{fig:wii10ka}~(b)).
The ratio of these two magnitudes is called the transformer ratio $R_w \equiv \left|E_z^w/E_z^b\right|$.
The energy gain of a witness electron beam, placed at the rear position of the wake is thus given by
$\Delta \gamma_w \, m c^2 = R_w~\gamma_b \, m c^2$ after energy depletion of an electron driver beam with initial mean Lorentz factor of $\gamma_b$.
Longitudinally symmetric drivers cannot excite plasma waves with transformer ratios $R_w > 2$ in the linear regime~\cite{Chen:1986sa}.
However, in the blowout regime, symmetric Gaussian profiles can generate transformer ratios $R_w > 3$ (cf.~Figure~\ref{fig:wii10ka}(b)),  
and triangularly ramped profiles~\cite{lotov:053105} can reach even $R_w > 5$.
The Wakefield-Induced Ionization (WII) injection method exploits the fact that wakefields in the blowout regime in PWFA have transformer ratios significantly greater than one to induce ionization and trapping 
of high-quality electron bunches into the extreme accelerating fields of the plasma wake.
These electrons originate from an atomic species with high ionization threshold (HIT) 
which is doped into the background plasma in a short axial region of the plasma target.

\begin{figure}[!h]
 \centering
  \includegraphics[width=1.0\columnwidth]{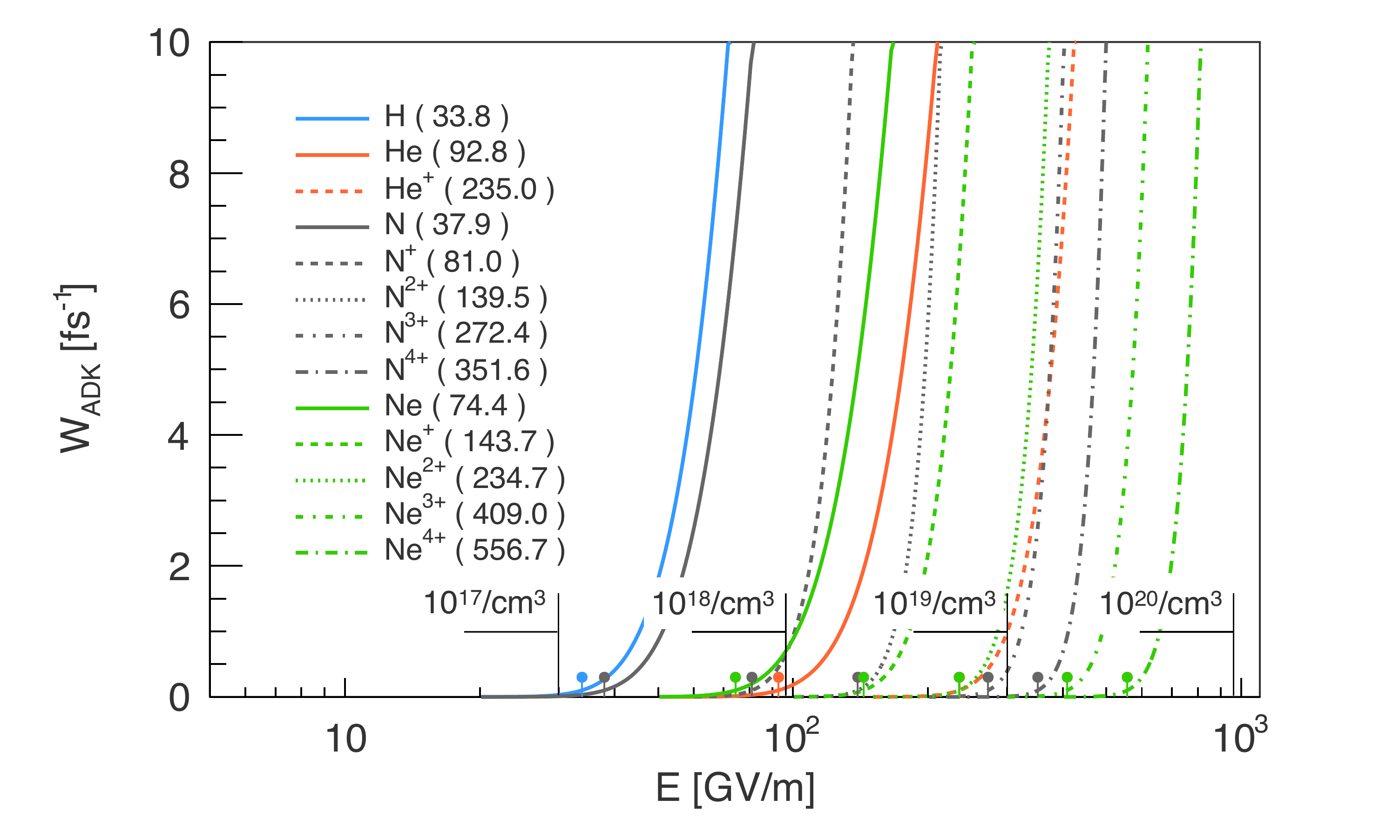}
  \caption{
    Ionization probability rates according to the theory by Ammosov, Delone, and Krainov (ADK)~\cite{Ammosov:1986adk}.
    The ionization thresholds are indicated by the dots and their numerical values written in units of $\mathrm{GV/m}$.
    The flags on the x-axis are placed at field values corresponding to $E_0$ for four different plasma densities.
   }
\label{fig:adkprob} 
\end{figure}

The ionization process caused by static (or slowly varying compared to the ionization process) electric fields 
of a magnitude sufficient to significantly deform the atomic potential barrier 
can be described by a tunnelling probability~\cite{perlomov:66}, 
which has been determined for a number of atomic species~\cite{Ammosov:1986adk}. 
In Figure~\ref{fig:adkprob} the ionization probability rates ($W_{\mathrm{I}}$) are depicted
as a function of the total electric field $E \equiv |\vec{E}|$, for some different types of gases.
These probability rates grow exponentially as soon as $E$
approaches an ionization threshold, which is defined here as     
the field $E_\mathrm{ion}$ for which the ionization probability rate becomes $W_{\mathrm{I}}=0.1~\mathrm{fs^{-1}}$.
According to this definition, the ionization threshold is written in units of $\mathrm{GV/m}$ 
in Figure~\ref{fig:adkprob} for the set of selected atomic species.

Generally, the field configuration of the beam driver and wakefield in the blowout regime
enables both ionization and trapping from two regions within the first wave 
bucket~\cite{Kirby:2009zzb}.
One region is located at the driver beam position, where the combination of
the radial electric field of the beam driver and the electric wakefield can induce ionization, if sufficiently high.
In the following the maximum electric field magnitude in the region of a high current, tightly focused beam driver is computed to estimate the conditions for the onset of ionization in this region.
The radial electric field in the blowout can be calculated via Gauss' law, for a given charge density of the drive beam and with the homogeneous ion charge density in the blowout.
For axially symmetric Gaussian beams ($n_b=n_b^0\,e^{-\zeta^2/2\sigma_z^2-r^2/2\sigma_r^2}$),
the radial electric field is then given by
\begin{equation}
\frac{E_r}{E_0} = - \frac{\Lambda_b}{k_pr}~\left(1-e^{-\frac{r^2}{2\sigma_r^2}}\right)+\left(1- \frac{\partial_\zeta E_z}{k_pE_0}\right)~\frac{k_pr}{2}\,,\label{eq:erbeamion}
\end{equation}
where $\Lambda_b(\zeta) = k_p^2 \int_0^\infty \frac{n_b}{n_0} r\mathrm{d}r$
is the normalized charge per unit length of the beam.
$\Lambda_b$ is related to the electric current of the beam by 
$\Lambda_b = (k_p^2 / 2 \pi e c n_0)~I_b = 2~I_b/I_A$, 
where $I_A = 4\pi \epsilon_0 mc^3/e \simeq 17.05~\mathrm{kA} $
is the Alfv\`en current.
The above defined beam thus has the normalized current profile $\Lambda_b= (n_b^0/n_0)~(k_p\sigma_r)^2~e^{-\zeta^2/2\sigma_z^2}$.
The first term in Eq.~(\ref{eq:erbeamion}) corresponds to the contribution from the beam charge density,
while the second term represents the contribution from the uniform charge density of ions
plus the slope of the longitudinal wakefield inside the cavity.
In the drive beam region, $k_p^{-1} \partial_\zeta E_z/E_0 \ll 1$ and this contribution to the second term in Eq.~(\ref{eq:erbeamion}) can be neglected.
Furthermore, for high-current ($\Lambda_b \gtrsim 1$), tightly focused ($k_p \sigma_r \lesssim 1$) beams, the first term in this equation dominates over the second term and the maximum of $E_r$ can be obtained analytically from Eq.~(\ref{eq:erbeamion})
\begin{equation}
\frac{E_r^{\mathrm{max}}}{E_0} \simeq -0.45~\frac{\Lambda_b}{k_p\sigma_r}\,.\label{eq:erbeammax}
\end{equation}
This maximum value is reached at distance $r \simeq 1.585~\sigma_r$ from the center of the beam's propagation axis.
Given this inverse proportionality of $E_r^{\mathrm{max}}$ on $\sigma_r$,
the magnitude of the electric field in this region is highly sensitive
to the betatron oscillations of the beam envelope $\sigma_r$ in the focusing ion-cavity~\cite{clayton:2002,Esarey:2002beta}
and fluctuations in the slice-emittance and current profile.

The second region from which ionization and subsequent trapping can occur is located at the rear of the ion-cavity, where only the wakefields induce ionization.
While the electric field magnitude in the beam region depends on details of the beam's charge distribution,
the electric wakefield behind the driver, in the back of the cavity,
does not depend on these details, but only on the geometrical structure of the blowout.
An important feature of the blowout regime may be noted in this regard:
The structure of the plasma wake is mainly sensitive to the
integrated charge per unit length of the driver, and not to its transverse distribution,
as long as the envelope of the drive beam is sufficiently smaller than the blowout radius $\sigma_r \ll r_l$.
This can be seen in Eq.~(\ref{eq:erbeamion}), where the impact of the electric field
on the plasma electrons in the sheath for $r=r_l \gg \sigma_r$ is dominated
by the normalized current profile of the drive beam $\Lambda_b$.
This implies that the wakefields are not sensitive to betatron oscillations
and fluctuations of the driver beam distribution, and remain stable.
The WII injection exploits this fact to induce ionization by means of
the wakefields only, while avoiding any contribution from the electric fields in the driver's region. 
For this reason, the dopant species has to be chosen to have an ionization field threshold
above the maximum value in the driver's region $E^b$, 
and smaller than the electric field magnitude at the back of the cavity $E^w$.
For high-current, moderately wide beams, i.e.~$k_p\sigma_r \approx 1$,
and $\sigma_r < r_l$, the radial electric beam field is 
significantly smaller than the longitudinal magnitude of the wakefield in regions near the propagation axis.
Given the absence of any current, also the magnitude of the electric field in the back of the cavity behind the drive beam
can be approximated by the longitudinal component only.
The condition for ionization to occur only in the rear part of the cavity
might hence be written as 
\begin{equation}
|E_z^w| > E_{\mathrm{ion}} > |E_z^b| \,, \label{eq:ionth}
\end{equation}
with $E_{\mathrm{ion}}$ the ionization threshold of the given HIT species.
In order to complete the injection, ionization must be followed by trapping of the released electrons.
Electrons are only trapped if they do not escape the focussing channel of the blowout
while gaining a sufficiently large velocity to co-propagate with the plasma wake at $v_b\approx c$.
Considering that the electrons are released at rest in a phase position $\zeta_i$, 
the velocity of the ultrarelativistic wake can be only acquired if the electrons end up in a phase position $\zeta_f$ that satisfies
\begin{equation}
\Delta\psi \equiv \psi_i-\psi_f = 1\,, \label{eq:trapping}
\end{equation}
where $\psi$ is the normalized wake potential, related to the wakefields by
$E_z/E_0 = -\partial_\zeta \psi /k_p$ and $W_\perp/E_0 = -\partial_r \psi /k_p$.
Eq.~(\ref{eq:trapping}) is a necessary but not sufficient condition for trapping,
which is obtained from the constant of motion of electrons~\cite{Oz:2007mn} under the quasi-static approximation~\cite{sprangle1990}.
Thus, in order to allow trapping from ionization, 
the plasma wake must provide a minimum difference in normalized wake potential of $\Delta \psi = 1$. 
Figure~\ref{fig:wii10ka}~(e) shows the structure of $\psi$ in the central slice of the example simulation.
The maximum potential $\psi_m$ is reached at the center of the blowout cavity ($\zeta_m$), 
where $E_z$ changes from positive to negative and the cavity radius is maximal ($r_l=r_m$). 
The minimum $\psi_{\mathrm{min}}$ is located at the very end of the cavity, $\zeta_c$, 
where the plasma electrons cross the propagation axis and $E_z$ changes from negative to positive values (cf.~Figure~\ref{fig:wii10ka}(e)).
The maximum potential difference of $\psi$ is therefore given by the integral 
$\Delta \psi_{\mathrm{max}} = -k_p \int_{\zeta_m}^{\zeta_c} (E_z/E_0)~\mathrm{d}\zeta$.
The phenomenological models for the blowout regime~\cite{Lotov:2004zz,Lu.PRL.96.165002}
connect $E_z$ with the dynamics of the plasma electrons in the plasma sheath.
A particularly clear differential equation for $E_z$ is provided in~\cite{Lotov:2004zz},
\begin{equation}
\frac{\partial_\zeta E_z}{k_p E_0} = - \frac{2\Lambda_b}{(k_p r_l)^2} + \frac{4}{(k_p r_l)^2} \left[\frac{E_z}{E_0}\right]^2 - \frac{\beta_{lz}}{1-\beta_{lz}}\,. \label{eq:boeqE}
\end{equation}
Eq.~(\ref{eq:boeqE}) explicitly expresses the dependence of $\partial_\zeta E_z$ on the cavity radius $r_l$,
the magnitude of $E_z$ itself, and finally, the longitudinal velocity $\beta_{lz}$
of the plasma electrons in the screening layer, and it is used for estimations in the context of this paper.
For sufficiently high-current Gaussian beams ending before the cavity center,
this model~(Eq.~(\ref{eq:boeqE})) connects both the maximum decelerating field in the region of the drive beam, $E_z^b=E_z(\zeta_b)$,
and the maximum blowout radius, $r_m$, with the peak current of the drive beam, $\Lambda_b^0$,
\begin{subequations}
\begin{align}
E_z^b/E_0 &\approx \sqrt{\Lambda_b^0/2}\,, \label{eq:ezbmax} \\
(k_pr_m)^2 &\approx \sqrt{32\pi\Lambda_b^0}~(k_p\sigma_z)\,. \label{eq:radiuslo} 
\end{align}
\end{subequations}
In addition, it can be seen that, in case of a large blowout radius $k_pr_m\gg1$,
the electrons in the plasma sheath acquire high speed in backwards direction
in respect to the driver beam propagation.
If the beam has a negligible current at the region of maximum radius within the blowout, 
one finds from Eq.~(\ref{eq:boeqE}) that 
$E_z$ is proportional to $\zeta$ in a wide region from the cavity center towards the back.
The slope of the longitudinal field is defined entirely by the plasma sheath velocity
\begin{equation}
\frac{\partial_\zeta E_z}{k_p E_0} = - \frac{\beta_{lz}}{1-\beta_{lz}} \approx \frac{1}{2}\,, \label{eq:ezcenter}
\end{equation}
where the last approximation is obtained in the limit of $\beta_{lz} \rightarrow -1$.
Under the assumption that $E_z$ continues with this linear slope from $\zeta_m$ up to the end of the cavity
$\zeta_c$,
it is possible to find a simple estimate for the magnitude of $E_z^w$,
and the maximum difference in wake potential
\begin{subequations}
\begin{align}
E_z^w/E_0 &\approx \frac{k_pl_m}{2}~\label{eq:ezmax}\,, \\
\Delta \psi_{\mathrm{max}} &\approx \left(\frac{k_pl_m}{2}\right)^2\,, \label{eq:lotovradius} 
\end{align}
\end{subequations}
where $l_m\equiv\zeta_m-\zeta_c$ is defined as the longitudinal semi-axis of the ion-cavity (Figure~\ref{fig:wii10ka}~(a)).
\begin{figure}[!t]
 \centering
  \includegraphics[width=1.0\columnwidth]{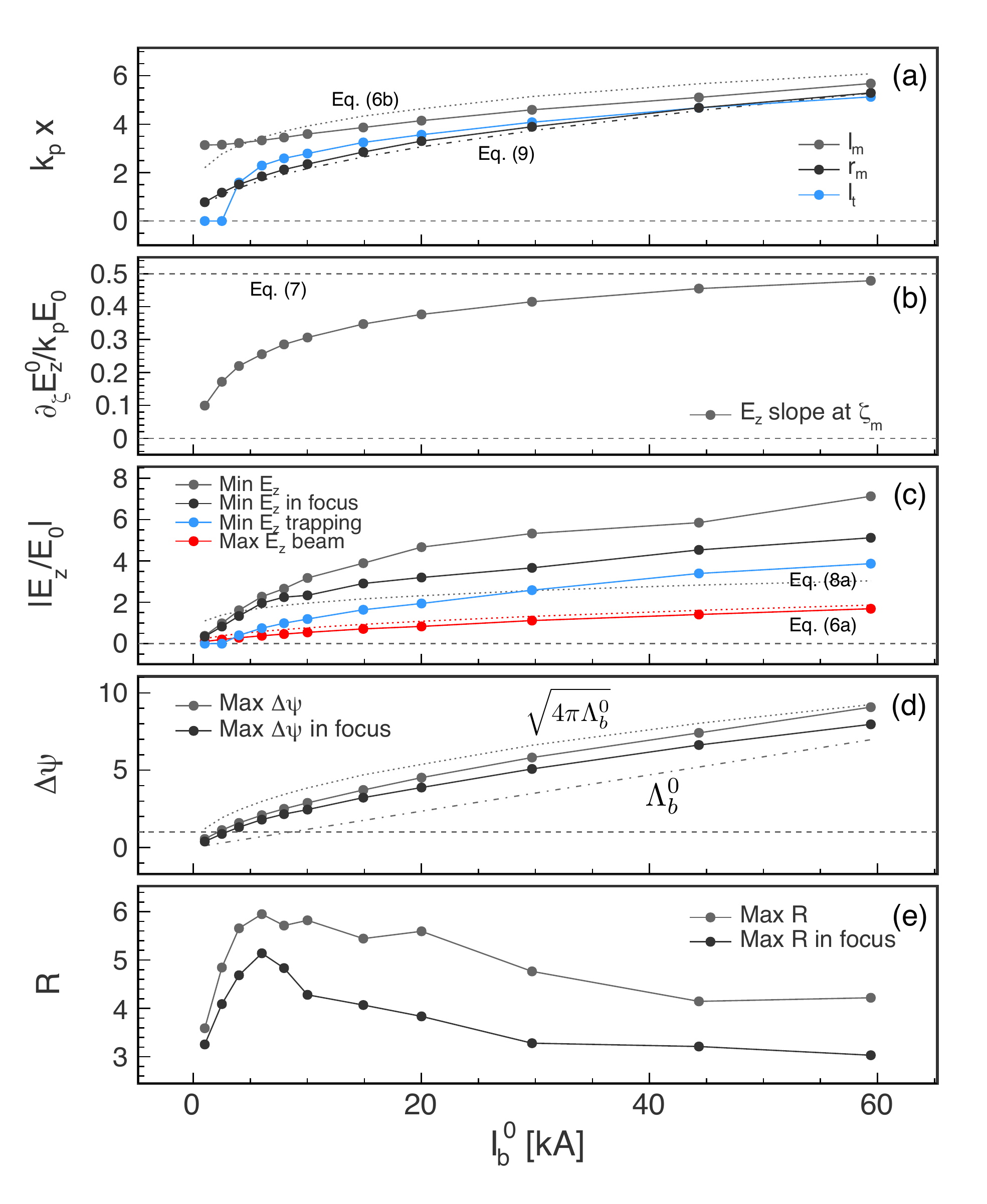}
  \caption{
    Scaling of key parameters in the blowout regime of PWFA as a function of the peak current of Gaussian drive beams.
    (a) Transverse ($r_m$) and longitudinal ($l_m$) semi-axis of the ion cavity plus the semi-axis of the trapping zone ($l_t$).
    (b) Derivative of $E_z$ with respect to $\zeta$ at the cavity center. (c) Magnitude of the maximum $E_z$
    at the back end of the cavity (grey line), inside of the focusing region (black line), in the trapping zone (blue line) and in the driver region (red line). 
    (d) Maximum wake potential difference in the blowout cavity (grey line) and in the focusing region (black line).
    (e) Maximum transformer ratio in the cavity (grey line) and in the focusing region (black line). The dotted lines represent certain model estimations that are explained on the text.
  }
\label{fig:boutscal} 
\end{figure}

In this work, we compare these approximate scalings with results from 3D simulations obtained with the PIC code OSIRIS.
The series of simulations consists on axially symmetric Gaussian beams of fixed geometrical dimensions ($k_p\sigma_z = \sqrt{2}$ and $k_p\sigma_r=0.1$),
fixed normalized emittance ($k_p \epsilon_n = 0.31$) and variable peak current $I_b^0$, ranging from $1~\mathrm{kA}$ up to $60~\mathrm{kA}$.
Figure~\ref{fig:boutscal}(a) shows $r_m$ (black line) and $l_m$ (grey line) as a function
of the peak current of the beam.
The top dashed line is the blowout radius obtained from Eq.~(\ref{eq:radiuslo}) for $k_p\sigma_z = \sqrt{2}$, i.e. $8\sqrt{\pi\Lambda_b^0}$.
It is worth noting that, despite the fact that this radius overestimates $r_m$ for relatively low currents, 
it reproduces the scaling of $l_m$.
The second dotted line is defined by 
\begin{equation}
k_pr_m \approx 2 \sqrt{\Lambda_b^0}\,,  \label{eq:radiuslu}
\end{equation}
which is a well known scaling of the transverse blowout radius as a function of the driver current, 
for moderate length Gaussian drivers ($k_p\sigma_z \approx \sqrt{2}$), obtained from PIC simulations~\cite{clayton:2002,Lu.PRL.96.165002}.
For high peak currents, $r_m$ obtained from Eq.~(\ref{eq:radiuslu}) as well as
$r_m$ obtained from Eq.~(\ref{eq:radiuslo}) asymptotically converge towards $l_m$ in the PIC simulations. 
The sheath of plasma electrons hence aproximately follows a spherical shape behind the driver beam for high current drivers.
The scaling for the blowout radius with the beam peak current in Eq.~(\ref{eq:radiuslo}) in this comparison is identified as an accurate estimation for $l_m$ (cf.~Figure~\ref{fig:boutscal}(a)).
Hence, the maximum $\psi$ difference may be connected with the current of the driver beam 
$\Delta\psi_{\mathrm{max}} \approx \sqrt{4\pi \Lambda_b^0}$, by means of this scaling.
In contrast, the scaling for $r_m$ from Eq.~(\ref{eq:radiuslu}) was used in a previous work~\cite{Kirby:2009zzb}
to express the maximum wake potential difference as a function of the driver's current $\Delta\psi_{\mathrm{max}} \approx \Lambda_b^0$,
and hence, estimate the minimum peak current for Gaussian beams to allow for trapping from ionization: $\Lambda_b^0>1$. 
In Figure~\ref{fig:boutscal}(d), we show the scaling of $\Delta\psi_{\mathrm{max}}$, 
as obtained from full 3D PIC simulations (grey line), together with the two estimations above. 
The dotted line shows $\sqrt{4\pi \Lambda_b^0}$, while the dash-dotted line depicts $\Lambda_b^0$.
Although slightly above, 
the $\sqrt{4\pi \Lambda_b^0}$ scaling seems to give a better description of the scaling of $\Delta\psi_{\mathrm{max}}$ from PIC simulations.

\begin{figure}[!t]
 \centering
  \includegraphics[width=1.0\columnwidth]{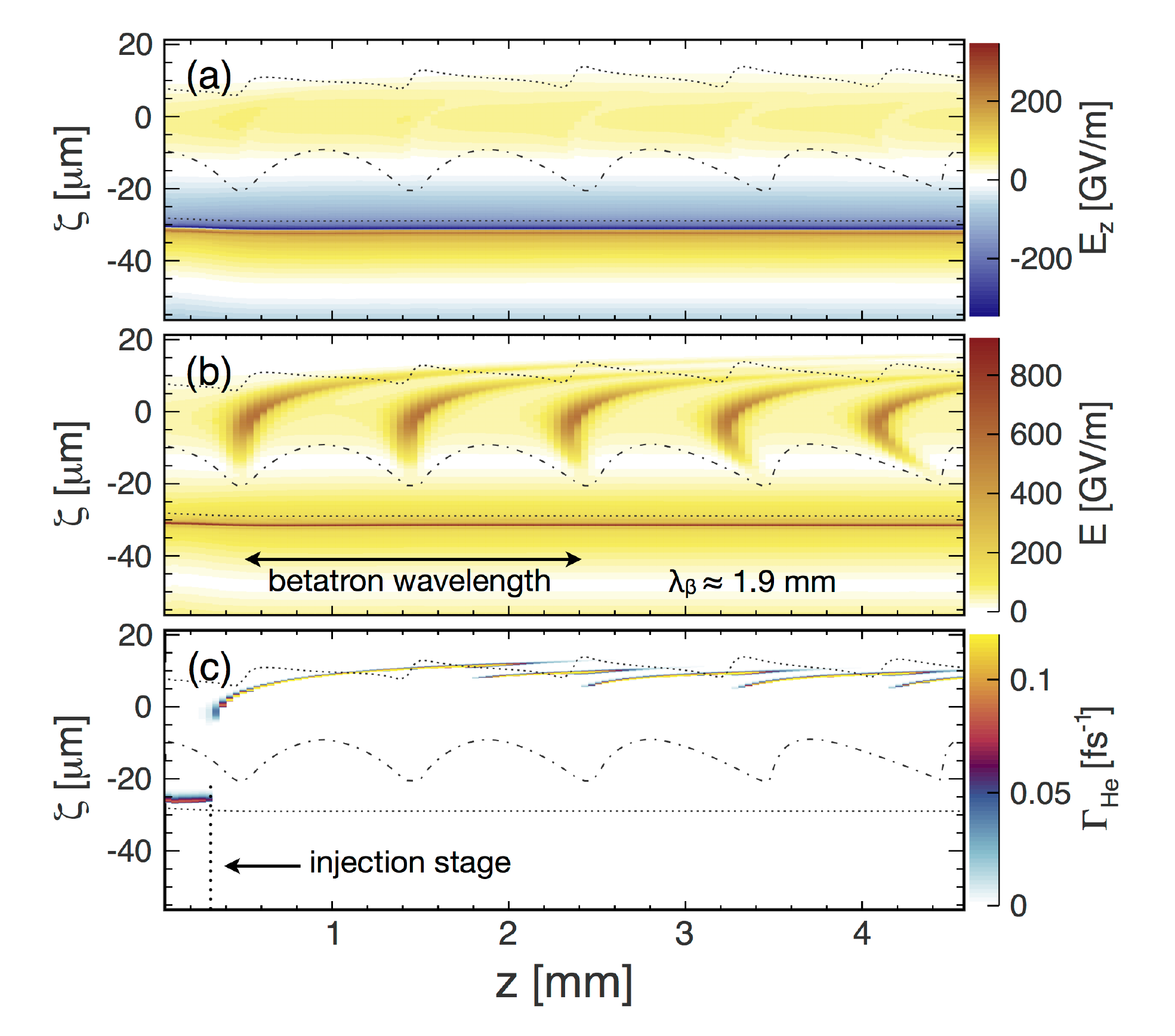}
  \caption{
    Evolution of the longitudinal electric field (a), the total electric field
    (b) and the ionization rate from He (c) in the on-axis region, 
    for an unmatched $10~\mathrm{kA}$ Gaussian electron driver.
    The dotted lines delimit the on-axis phase range with $\psi-\psi_F>1$.
    The dotted-dashed lines represent the end of the region where the radial fields of the beam are highly defocusing for slow electrons ($E_r\ll0$). 
  }
\label{fig:betaosc} 
\end{figure}

To be trapped and transported, electrons not only need to reach the velocity of the wake, but also must remain in the focusing region of the plasma wave. 
PIC simulations show that the position where the cavity closes and where $\psi$ is minimum falls in 
a defocusing region, so that electrons cannot be transported there.
The trapped orbits have to lie completely in the focussing region,
ahead of the co-moving position $\zeta_F$ at which the field configuration changes from focusing to defocusing (Figure~\ref{fig:wii10ka}(c)).
The black line in Figure~\ref{fig:boutscal}(d) shows the maximum potential difference that can be reached 
inside the focusing region, i.e. $\psi_m - \psi_F$, 
which is clearly smaller than the absolute $\Delta\psi_{\mathrm{max}}$ (grey line).
The maximum accelerating field in the focusing region is given by 
$E_z^F \equiv E_z(\zeta_F)$ (black line in Figure~\ref{fig:boutscal}(c)),
to be compared with the absolute minimum at the very end of the cavity (grey line).
To be trapped exactly at $\zeta_F$, an electron must be ionized at a position $\zeta_I>\zeta_F$, such that $\psi_I-\psi_F = 1$. 
The set of points fulfilling this condition form a surface that delimits the \emph{trappable zone}. 
Electrons released (at rest) inside this zone can \emph{potentially} be trapped 
in the focusing region of the blowout cavity. 
The blue line in Figure~\ref{fig:boutscal}(a) shows $l_t\equiv\zeta_m-\zeta_I$ (the \emph{trapping distance}) as a function of $I_b^0$, as obtained from the set of PIC simulations. 
We see that a \emph{trappable zone} ($l_t>0$) only exists for sufficiently high peak currents $I_b^0 \gtrsim 5~\mathrm{kA}$.
This defines a lower limit for the peak current of Gaussian drivers to achieve trapping of electrons based on ionization injection techniques in the quasi-static picture.
Nevertheless, we would like to point out here that this limit can be lowered if the phase velocity of the wake
is reduced when performing the ionization in a plasma density transition.
In order to induce ionization in the trappable region by means of the wakefields at the back of the blowout, 
the magnitude of the accelerating  field at $\zeta_I$, $E_z^I \equiv E_z(\zeta_I)$
(blue line in Figure~\ref{fig:boutscal}(c)) 
has to be greater than the ionization threshold of the dopant species $|E_z^I| > E_\mathrm{ion}$.
In addition, it is important to avoid any spurious contribution from the evolving electric fields in the driver's region (cf.~Figure~\ref{fig:betaosc}(b)).  
As we saw from Eq.~(\ref{eq:erbeamion}),
the radial electric field is sensitive to the envelope betatron oscillations of the beam,
and it increases significantly when the beam envelope is reduced.
Electrons ionized by the radial fields in the driver beam region gain transverse momentum
while they are still slow ($\ll c$), and are likely to escape the relativistic focusing cavity in the transverse direction. 
Another reason to avoid ionization in the beam region is the fact that narrow beams can ionize entirely the dopant species around the axis,
leaving no atoms or molecules which can be ionized in the rear wakefields.   
For these reasons, beams in WII injection should not trigger ionization themselves in regions near the axis.
It is interesting to note that the beam spot size that makes $E_r^{\mathrm{max}}$ in Eq.~(\ref{eq:erbeammax})
equal to $E_z^b$ in Eq.~(\ref{eq:ezbmax}) is given by $k_p\sigma_r \simeq 0.64~\sqrt{\Lambda_b^0}$,
which is always much smaller than the blowout radius in Eq.~(\ref{eq:radiuslu}).
The latter condition ensures that the total electric field is dominated by its longitudinal component
in the beams region near the propagation axis, while mantaining the condition for a stable blowout ($\sigma_r<r_m$).
However, unless the beam is matched to the focusing plasma channel, this situation does not last,
and the injection must occur before the beam is transversely compressed due to the betatron oscillations within the ion channel.
After the injection, it is preferable that the driver beam is tightly focused 
to ensure a stable propagation that mitigates the head erosion~\cite{An:PhysRevSTAB13}.
This sets a condition for the normalized transverse emittance of the beam
to be smaller than the matched value $k_p\epsilon_n < \sqrt{\gamma/2}~(k_p\sigma_r)^2$.
The spatial period of the envelope oscillations of the beam
is given by half the betatron wavelength~\cite{Esarey:2002beta}
of the beam-particles in the blowout $\lambda_{\beta}=\sqrt{2\gamma}/k_p$.
Figure~\ref{fig:betaosc} shows the evolution of $E_z$ (a)
and $E=|\vec{E}|$ (b) in the on-axis region ($k_pr<0.1$),
for the Gaussian driver beam in the example of Figure~\ref{fig:wii10ka}.
Due to the envelope oscillations of the beam, the total electric field in the driver's region oscillates with a spatial periodicity of $\lambda_\beta/2$ (Figure~\ref{fig:betaosc}(b)).
On the contrary, the longitudinal fields (in Figure~\ref{fig:betaosc}(a)) are stable during the propagation.
After the beam is transversely compressed,
the region with a high ionization rate moves from the back of the cavity to the front of the beam (Figure~\ref{fig:betaosc}(c)).

Figure~\ref{fig:boutscal}(c) depicts the on-axis values for the longitudinal electric field at the smallest possible co-moving position from which electrons can still be trapped $E_z^I=E_z(\zeta_I)$ (blue line) and the maximum longitudinal electric field in the beam region $E_z^b$ (red line).
The longitudinal electric field in the beam region according to Eq.~(\ref{eq:ezbmax}) is represented by the red dotted line, and closely follows the curve from the simulation data.
The grey dotted line in Figure~\ref{fig:boutscal}(c) is the result of extrapolating the 
linear tendency of $E_z$ at the cavity center (Eq.~(\ref{eq:ezcenter})) up to the end of the cavity $\zeta_c$. 
PIC simulations show that the difference in magnitude between $E_z^I$ and $E_z^b$ 
grows with increasing peak current of the driver,
hence allowing a wider range of ionization thresholds for which the ionization in the beam region is suppressed
but for which the ionization and trapping in the blowout is still possible.
The appropriate dopant species for WII injection must have
an ionization threshold $|E_z^I|> E_\mathrm{ion} > |E_z^b|$ (blue and red lines in Figure~\ref{fig:boutscal}(c)).
Such an ionization threshold can only be obtained using high-current beams $I_b^0 \gtrsim 8.5~\mathrm{kA}$.
The criterion for the current corresponds to a normalized current of $\Lambda_b^0 \simeq 1$, and is independent of the plasma density.
In addition, Figure~\ref{fig:boutscal}(b) depicts
the slope of $E_z$ around the cavity center (grey line),
which converges to the value suggested by Eq.~(\ref{eq:ezcenter}) for high-current drivers.
Figure~\ref{fig:boutscal}(e) shows the overall maximum transformer ratio (grey line) and
the maximum transformer ratio for witness beams placed within the focusing region (black line) in the PIC simulations.
\begin{figure}[!t]
 \centering
  \includegraphics[width=1.0\columnwidth]{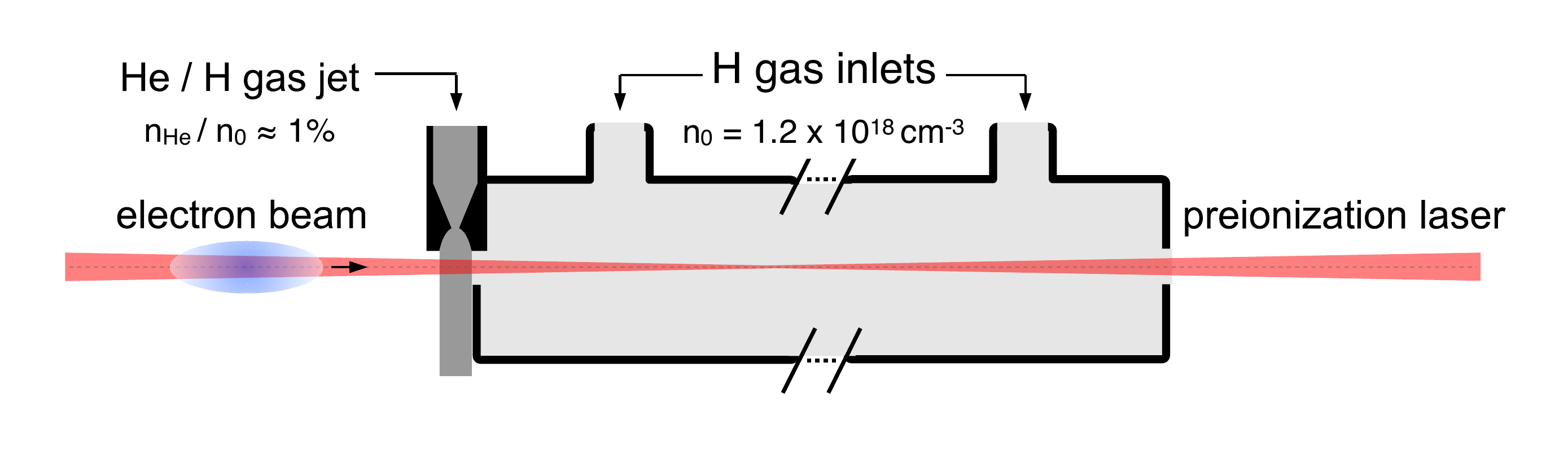}
  \caption{
 Schematic of the plasma-cell setup assumed as a basis for the OSIRIS 3D simulations. 
 A thin jet of a neutral hydrogen/helium gas mixture upstream of a flat-top hydrogen gas target assures localized WII injection. The hydrogen is pre-ionized with a laser while the helium remains non-ionized.
  }
\label{fig:plasmacell} 
\end{figure}

The theoretical framework for WII injection described above is independent of the choice for the plasma density. 
However, the choice of a specific plasma density with a specific $E_0$ requires the choice of an appropriate dopant species with suitable ionization threshold(s) (Figure~\ref{fig:adkprob}).
Ideally, the plasma density is chosen such that the drive beam generates wakefields with the greatest possible transformer ratio.
This is the case at the so-called resonant length, which in the linear regime~\cite{Lulinear:2005} is given by $k_p\sigma_z=\sqrt{2}$. 
The horizontal lines in Figure~\ref{fig:wii10ka}(d) indicate 
the ionization thresholds for H, He, and Ne in case of $E_0 = 105~\mathrm{GV/m}$, 
which corresponds to a plasma density of $n_0 = 1.2\times10^{18}~\mathrm{cm^{-3}}$.
For this particular example, the first ionization thresholds of He and Ne (solid orange and green lines in Figure~\ref{fig:wii10ka}(d), respectively) 
fulfil the condition expressed in Eq.~(\ref{eq:ionth}) and can be used for WII injection.

In order to exemplify and further analyze the characteristics of WII injection and the trapped bunches, results from PIC simulations on two experimental setups will be presented in the following.
These are the FLASHForward project at DESY and the FACET facility at SLAC. 
The experimental setup considered is sketched in Figure~\ref{fig:plasmacell};
The plasma is pre-created by a laser pulse with an intensity sufficiently high to fully ionize a gas with a low ionization threshold (LIT),
e.g.~hydrogen along the propagation axis of the drive beam.
In addition, a micro-nozzle~\cite{Hao:2005} fed by the same LIT gas doped with  
a tunable concentration of a high-ionization-threshold (HIT) gas (e.g.~helium),
is positioned in the vicinity of the gas-cell entrance. 
Because the electrons to be injected in the wake by means of Wakefield-Induced Ionization 
will originate from the HIT dopant gas, 
it must remain non-ionized after the passage of the ionization laser and the electron driver beam.
This technology for the plasma cell~\cite{Schaper2014208} also allows 
the density profile of the gas components to be controlled by means of several gas inlets with tunable pressure.
The density profile configuration for WII injection can be simple and we only consider the case in which 
the density of the LIT gas is approximately uniform all along the main channel and the micro-nozzle.
The doped gas jet emerging from the nozzle can spatially be confined to a distance which is significantly shorter than the betatron wavelength,
so that an unmatched beam experiences maximum transverse compression downstream of the doped gas nozzle.

\section{FLASHForward at DESY}
\label{sec:FLASHForward}
FLASHForward (Future-oriented wakefield-accelerator research and development at FLASH) is a plasma-wakefield acceleration project at DESY, 
which utilizes high current and short electron beams from the versatile FLASH accelerator.
The standard operational mode of the FLASH accelerator offers
electron bunches with approximately Gaussian longitudinal 
($\sigma_z \sim 20~\mathrm{\mu m}$) and radial ($\sigma_r = 10~\mathrm{\mu m}$) 
profiles, transverse normalized emittances of $\epsilon_n \approx 1~\mathrm{\mu m}$,
and an energy of $1~\mathrm{GeV}$ with a relative spread of $0.1~\%$.
Such bunches are to be compressed even further in the beam-extraction arc upstream of the FLASHForward plasma cell, to reach peak currents of up to $I_b \approx 10~\mathrm{kA}$ 
in a more compact size ($\sigma_z \sim 7~\mathrm{\mu m}$ and $\sigma_r = 4~\mathrm{\mu m}$).
Operating this current-enhanced FLASHForward driver at the resonant length ($k_p\sigma_z = \sqrt{2}$) 
requires a plasma density of $n_0 = 1.2\times10^{18}~\mathrm{cm^{-3}}$.
This results in parameters for the electron driver and the density 
that correspond to the parameters used for the PIC simulation in Figure~\ref{fig:wii10ka} 
for a $10~\mathrm{kA}$ Gaussian driver.
As pointed out at the end of Sec.~\ref{sec:wiii}, at this plasma density the first ionization levels 
of both He and Ne match the condition for WII injection (Eq.~(\ref{eq:ionth})) with a $10~\mathrm{kA}$ beam.
\begin{figure}[!t]
 \centering
  \includegraphics[width=1.0\columnwidth]{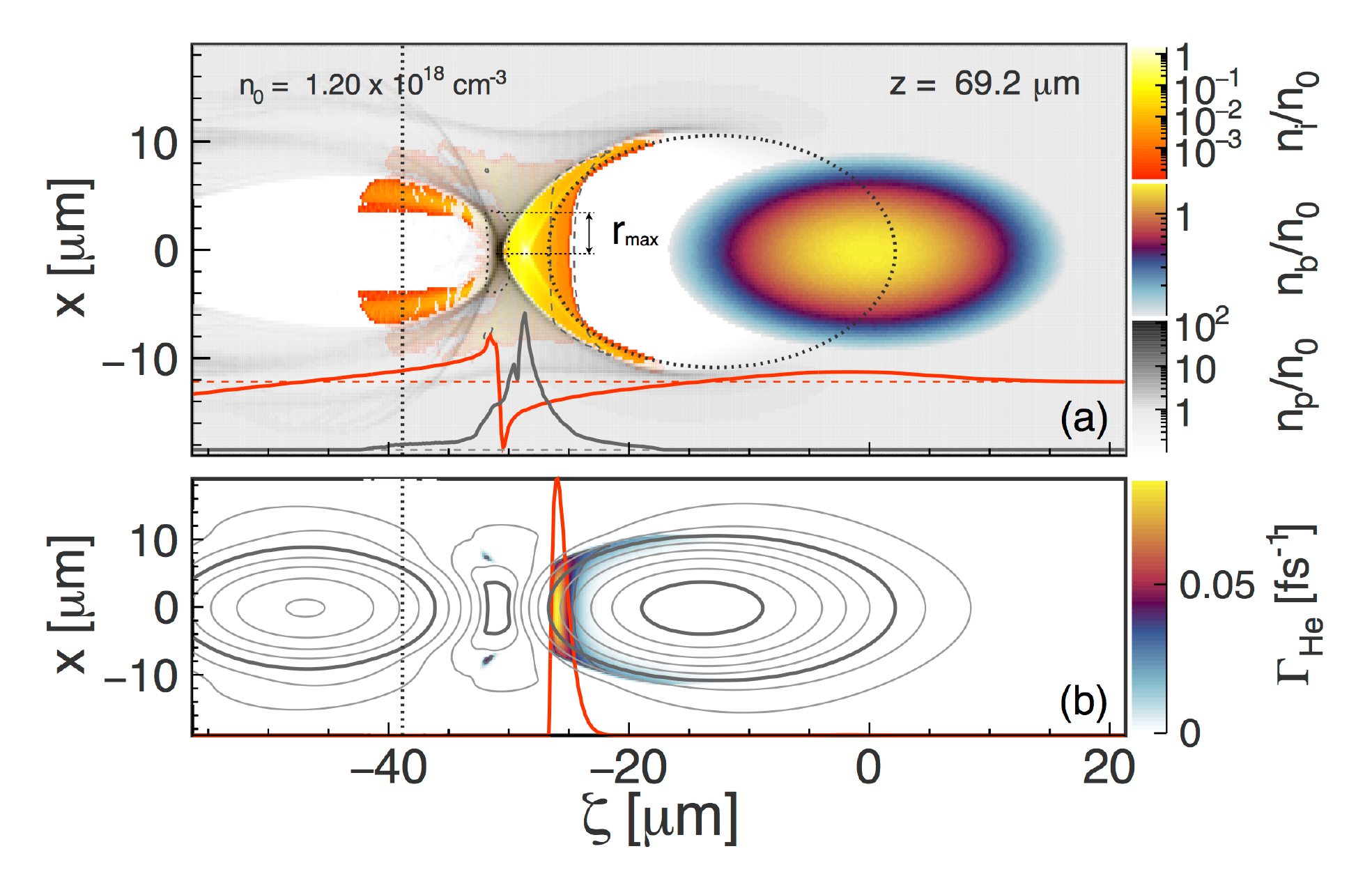}
  \caption{
    OSIRIS 3D simulation of a FLASHForward-type electron beam ($I_b = 10~\mathrm{kA}$),
    traversing a plasma at the resonant density ($n_0 = 1.2\times10^{18}~\mathrm{cm^{-3}}$). 
    (a) Spatial particle density. (b) Ionization rate for He according to the ADK model. 
    The contours in panel (b) show profiles of the equipotential surfaces. 
    These surfaces are depicted in steps of 0.2 starting from the minimum value inside the focusing region ($\psi_F$).   
    The vertical dotted line indicates the beginning of the He region.
  }
\label{fig:FFrake} 
\end{figure}
Figure~\ref{fig:FFrake}(a) shows a snapshot of the simulation when the beam
traverses a $L_{\mathrm{He}}=60~\mathrm{\mu m}$ diameter gas-jet doped with He at $n_{\mathrm{He}} = 0.01~n_0$ 
concentration.
Figure~\ref{fig:FFrake}(b) shows the expected ionization rate from the outer He electron
in a configuration with a $10~\mathrm{kA}$ peak current beam operating in a plasma of $n_0 = 1.2\times10^{18}~\mathrm{cm^{-3}}$.
Because the gas streams from right to left at $c$ in this co-moving picture,
the ionization rate is computed taking into account the He electrons that have been already ionized
\begin{equation}
\Gamma_{\mathrm{I}}(\zeta) \equiv W_{\mathrm{I}}(\zeta) (1-P_{\mathrm{I}}(\zeta)), 
\end{equation}
where $W_{\mathrm{I}}(\zeta) = W_{\mathrm{I}}(E(\zeta))$, the species ionization rate according to the ADK model~\cite{Ammosov:1986adk} 
and 
\begin{equation}
P_{\mathrm{I}}(\zeta) \equiv c^{-1} \int_\infty^{\zeta} W_{\mathrm{I}}(\zeta')~\mathrm{d}\zeta'\,,
\end{equation}
the ionization probability in the co-moving frame.
The regions with a high ionization rate will be denoted as the \emph{ionization volume}.
The volume from which injection is possible is thus determined by the intersection of the ionization volume
($E > E_\mathrm{ion}$) with the volume satisfying the trapping criterion ($\psi-\psi_F > 1$).
The contours of $\psi-\psi_F$ are drawn in Figure~\ref{fig:FFrake}(b), 
showing that only a narrow and well-defined region around the propagation axis 
satisfies the conditions for simultaneous ionization and trapping.
Owing to the small size of this injection volume, the trapped orbits will also be confined in a small transverse region and short phase interval during the acceleration process. 
These are prerequisites for the generation of intrinsically very short and low-emittance electron bunches. 
The ionization rate increases quickly as soon as the wakefield exceeds the ionization threshold,
thereby defining a narrow spike of high ionization rate in the on-axis region (red line in Figure~\ref{fig:FFrake}(b)).
Electrons originating from this location $\zeta_i$ are trapped at $\zeta_f$ complying with $\psi(\zeta_i)-\psi(\zeta_f) = 1$.
Because the $\psi$ contours are closer to each other when the electrons approach the trapped positions,
the longitudinal distribution during the acceleration of the injected electron bunch is narrower than the initial distribution.
In the considered example, the rms of the initial ionization volume is significantly smaller than the plasma skin depth $k_p^{-1}$.
Due to the invariance of $E_z$ along the cavity radius, the volume of injection extends transversely up to $r_l$.
Furthermore, the radial electric field increases linearly with increasing radius (Eq.~(\ref{eq:erbeamion})) and thus, 
the ionization rate is greater in positions close to the transverse boundary of the cavity.
However, electrons ionized close to the boundary are, either out of the trapping region 
or escape transversely due to sizeable transverse momentum gain from the positive transverse electric field in the most off-axis regions.
The initial injection volume may therefore be considered as a thin disc centred on axis 
which extends up to a radial position $r_{\mathrm{max}}$ that is always smaller than the cavity radius at that point (cf.~Figure~\ref{fig:FFrake}(a)).
The finally injected witness bunch is composed of electrons ionized from within this volume. 
Electrons belonging to the same final $\zeta_f$ slice originate from different radial positions 
along the initial $\psi_i$ contour.
Assuming full betatron decoherence for every slice, 
an upper estimate of the normalized transverse emittance
can be given in terms of the initial transverse extent of the slice~\cite{Kirby:2009zzb}
$\epsilon_n = k_p \langle r_i^2 \rangle/4$.
Considering for simplicity,
the ionization electrons to be uniformly distributed up to $r_{\mathrm{max}}$ along
the largest $\psi_i$ contour, the estimated maximum slice emittance yields 
\begin{equation}
k_p\epsilon_n^{\mathrm{max}} = \frac{(k_p r_{\mathrm{max}})^2}{12}\,.\label{eq:rakeemit}
\end{equation}
Typically $r_{\mathrm{max}}$ is much smaller that the maximum blowout radius $r_m$
provided that ionization and trapping only happens from the core of a thin ionization slice at a rear position of the blowout cavity.
For this reason, the WII injected bunches are, by construction, extremely compact and low emittance.
As a very practical rule of thumb, we can write the following relation for the WII injected bunches
\begin{equation}
k_p\sigma_z \sim k_p\epsilon_n \sim 0.1\,. \label{eq:wiibunchprop}
\end{equation}
That is, both the length and the normalized emittance of the injected bunch are a fraction of the plasma skin depth,
which scales with the inverse of the squared root of the plasma density ($k_p^{-1} \propto 1/\sqrt{n_0}$).
\begin{figure}[!t]
 \centering
  \includegraphics[width=1.0\columnwidth]{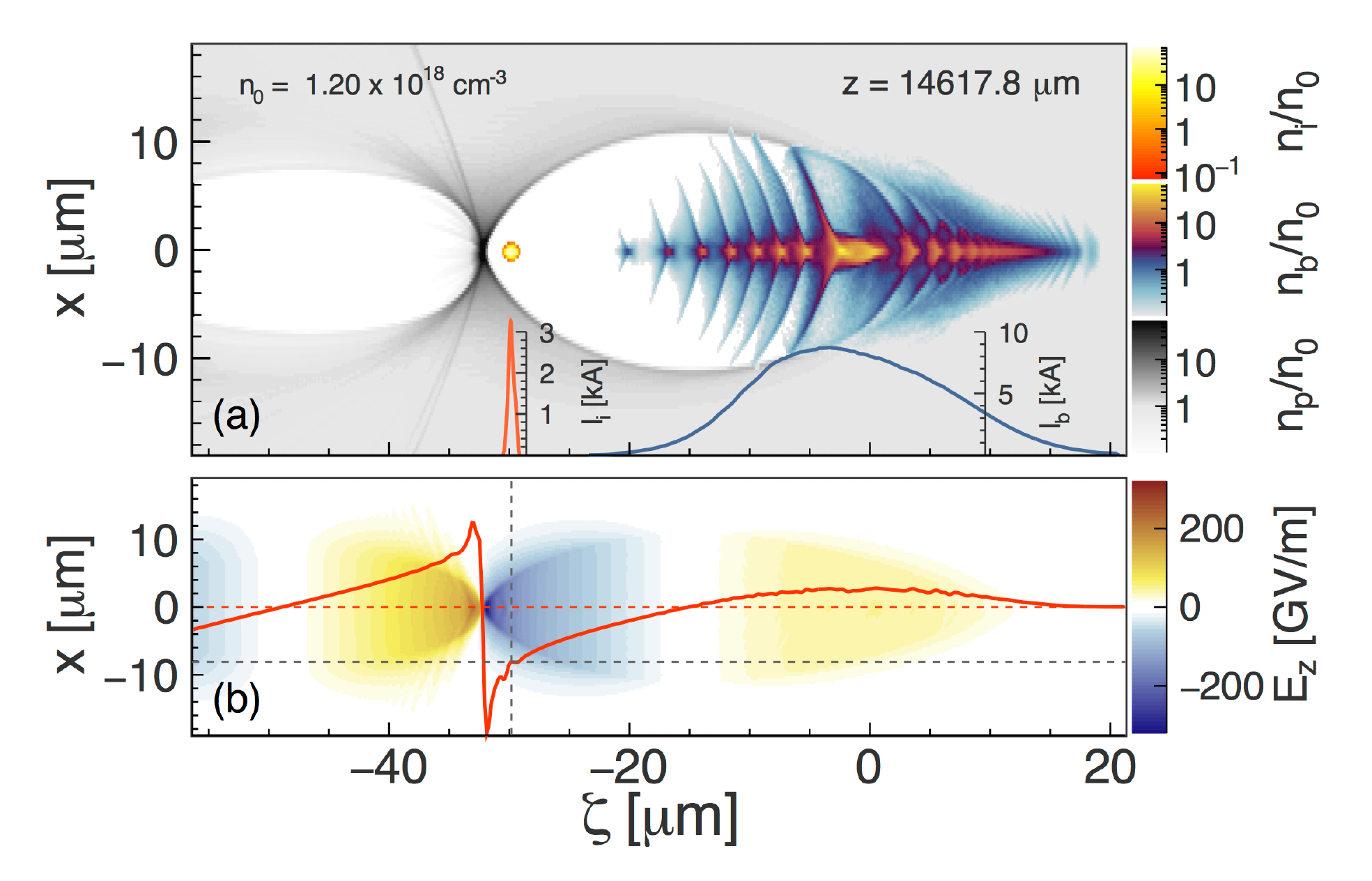}
  \caption{
    Results from the same PIC simulation as in Figure~\ref{fig:FFrake}, but after $14.6~\mathrm{mm}$ of propagation in plasma.
    (a) Spatial particle density. 
    Here a short bunch of $6.34~\mathrm{pC}$ has been injected and subsequently accelerated up to more than $2.5~\mathrm{GeV}$ energy.
    (b) Longitudinal electric fields. The bunch experiences an accelerating gradient substantially greater than $100~\mathrm{GV/m}$.   
  }
\label{fig:FFrakeprop} 
\end{figure}

Figure~\ref{fig:FFrakeprop} shows the density of the driver-witness system after $14.6~\mathrm{mm}$ 
of propagation in plasma in the simulation discussed above, 
where a short ($240~\mathrm{nm}$) bunch of electrons has been ionized and injected from the neutral He 
by means of the wakefields only.
The total injected charge amounts to $6.43~\mathrm{pC}$, 
while the rms current, defined as $I_{\mathrm{rms}} = c Q / (\sqrt{2\pi} \sigma_z)$, is equal to $3.2~\mathrm{kA}$.
Note that with this definition, the peak and rms currents for Gaussian beams are identical.
The injected beam has been accelerated over a distance of $14.6~\mathrm{mm}$,
from $\langle\zeta_f\rangle \approx -29.8~\mathrm{\mu m}$,
where $E_z(\langle\zeta_f\rangle) \approx 136~\mathrm{GV/m}$ (Figure~\ref{fig:FFrakeprop}(b)). 

\begin{figure}[!t]
 \centering
  \includegraphics[width=1.0\columnwidth]{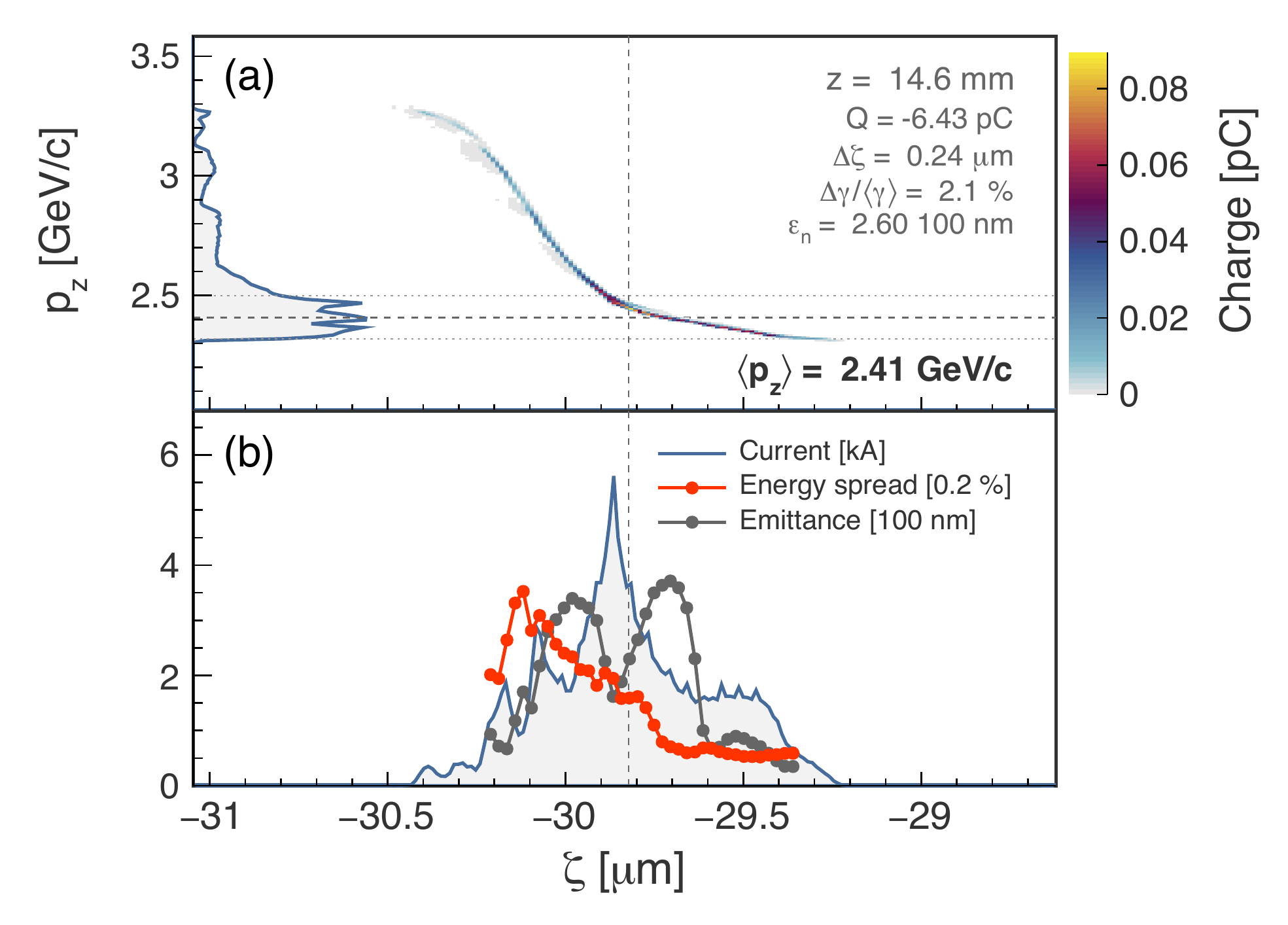}
  \caption{
    Witness bunch properties after $14.6~\mathrm{mm}$ of acceleration.
    (a) Absolute charge distribution of the bunch in longitudinal phase space ($p_{z}$ vs. $\zeta$ plane).
    The projection of this distribution in $p_{z}$ is depicted on the left axis.
    (b) Bunch-current profile.
    The relative energy spread and the transverse emittance 
    are plotted for different longitudinal slices along the bunch.   
  }
\label{fig:FFrakebunch} 
\end{figure}
Some additional properties of the trapped bunch can be estimated from the initial phase-space distribution as well.
Trapped electrons with the same initial $\psi_{i}$ propagate to the same final position $\zeta_f$, 
which fulfils $\psi_i - \psi_f = 1$; they will be accelerated by the same field value $E_{z,f}$. 
However, each one of these slices in $\zeta_{f}$ is composed of electrons ionized at different longitudinal
positions along the He column, and therefore are accelerated over different times, 
producing a finite spread in longitudinal momentum in every slice given by 
$\Delta p_{z,f} \simeq - e E_{z,f}~L_{\mathrm{He}}$,
which at the average position of the bunch gives 
$\Delta p_{z,f} \approx 8~\mathrm{MeV/c}$.
This is identified as the main contribution to the sliced energy spread of the beam, 
as other sources of thermal spread (e.g.~helium temperature) 
on the initital population of injected electrons are considered to be well below $\mathrm{MeV}$ levels.
Because this initial sliced energy spread due to the finite width of the dopant jet
is not expected to grow during the acceleration process,
the sliced relative energy spread of the bunch $\Delta\gamma_f/\langle\gamma_f\rangle = \Delta p_{z,f}/\langle\gamma_f\rangle mc$
can reach values well below $1\%$ at $1~\mathrm{GeV}$ electron energies.
On the other hand, the total relative energy spread is proportional to the variation of $E_z$ along
the bunch, which in case of negligible beam loading and sufficiently short bunches 
is approximately given by 
$\Delta\gamma/\langle\gamma\rangle \simeq (\partial_{\zeta}E_{z,f}/E_{z,f})~\sigma_{z,f}$.
This quantity does not depend either on the beam's electron energy or on the plasma density,
but only on the product of the local $E_z$ slope and its length.
Using Eq.~(\ref{eq:ezcenter}) to approximate the magnitude of the $E_z$ slope, and Eq.~(\ref{eq:rakeemit}) for the bunch size,
$\Delta\gamma/\langle\gamma\rangle \approx 5~\%$ for WII injected witness bunches.
From Figure~\ref{fig:FFrakeprop}(b), $\partial_{\zeta}E_z(\langle\zeta_f\rangle) \approx 28.9~\mathrm{(GV/m)\,\mu m^{-1}}$, 
and $\Delta\gamma/\langle\gamma\rangle \approx 8.8\mathrm{\%}$.

The properties of the simulated injected bunch after acceleration over a distance of $14.6~\mathrm{mm}$  
are summarized in Figure~\ref{fig:FFrakebunch}.
The longitudinal phase space (Figure~\ref{fig:FFrakebunch}(a)) exhibits a thin curved chirp 
with an average energy of $\sim2.55~\mathrm{GeV}$ and a projected relative energy spread of~$8.8\%$.
The horizontal dotted lines in Figure~\ref{fig:FFrakebunch}(a) delimit the full width at half maximum (FWHM)
for the projected energy spectrum on the left.
The value of relative energy spread ($\sim2~\%$) in FWHM is given in the figure.
Additional bunch properties can be seen in more detail in Figure~\ref{fig:FFrakebunch}(b).
The current profile has a maximum approximately at the center of the bunch of $\sim 5~\mathrm{kA}$.
The relative energy spread ($\lesssim0.3\%$), and the normalized transverse emittance ($\lesssim 300~\mathrm{nm}$)
are shown for different slices in $\zeta$.
The slice width ($0.005~\mathrm{k_p^{-1}}$) is chosen short enough such that a further reduction of this width
would not significantly change the values of the sliced properties, in order to appropriately reflect their uncorrelated values.
The maximum uncorrelated normalized emittance along the bunch is $k_p\epsilon_{n}^{\mathrm{max}} = 0.06$,
which may be connected with the injection volume maximal radius as discussed above.
Knowing the emittance in the simulation, the maximum ionization radius can be estimated by using  
Eq.~(\ref{eq:rakeemit}), $r_{\mathrm{max}} \approx 0.86~k_p^{-1} = 4.2~\mathrm{\mu m}$, 
which matches well with the simulations (cf. Figure~\ref{fig:FFrake}(a)). 

\section{FACET at SLAC}
\label{sec:FACET}
\begin{figure}[!t]
 \centering
  \includegraphics[width=1.0\columnwidth]{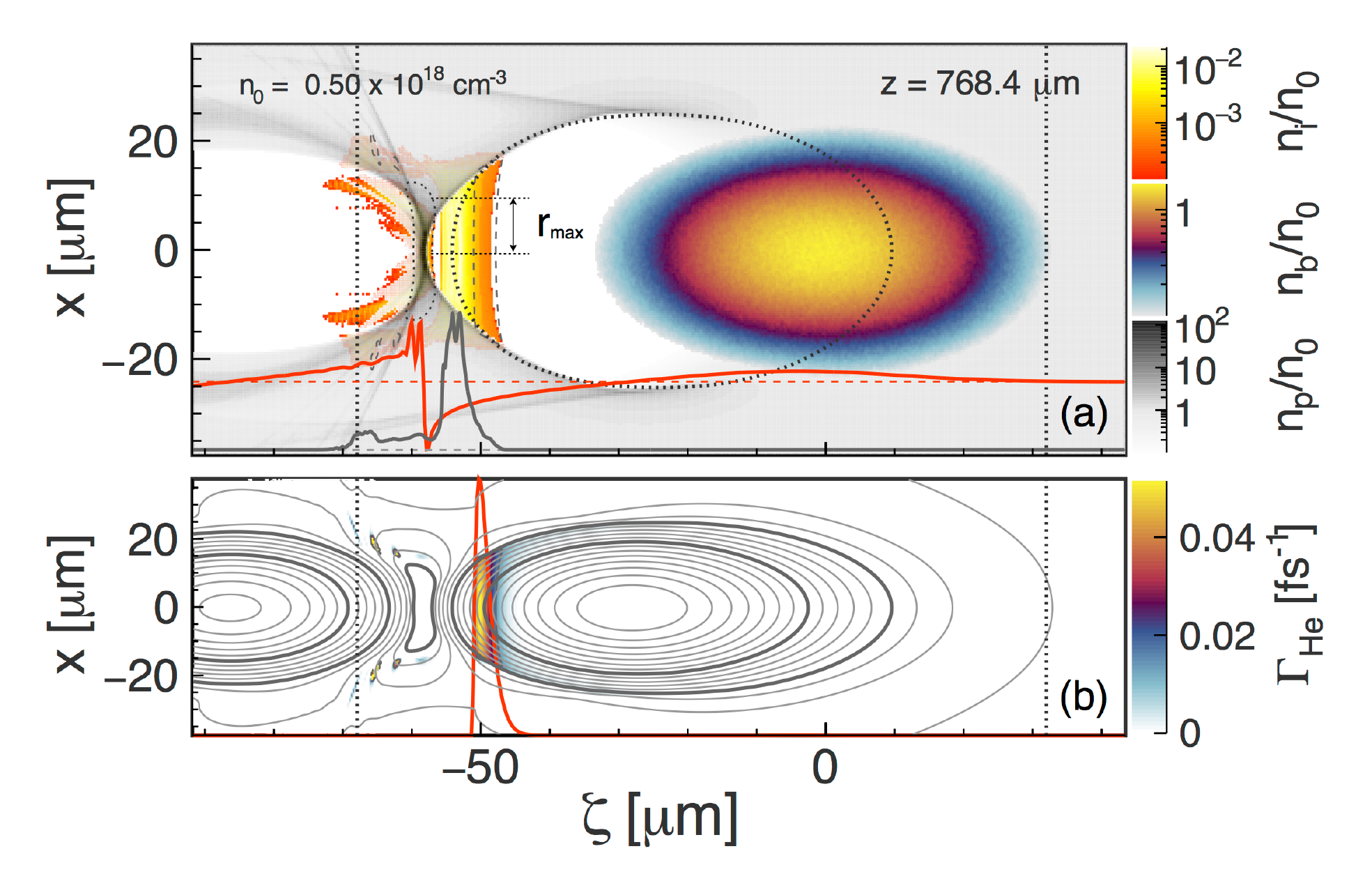}
  \caption{
    OSIRIS 3D simulation of a FACET electron beam ($I_b = 23~\mathrm{kA}$),
    traversing a plasma with $n_0 = 0.5\times10^{18}~\mathrm{cm^{-3}}$ and resonantly exciting a plasma wave. 
    (a) Spatial particle density. (b) Ionization rate map for He according with the ADK model. 
    The contours in panel (b) outline the iso-potential regions. 
    They are drawn in steps of 0.2 starting from the minimum value inside the focusing region ($\psi_F$).   
    The vertical dotted line indicates the beginning of the He region.
  }
\label{fig:FACETrake} 
\end{figure}

\begin{figure}[!t]
 \centering
  \includegraphics[width=1.0\columnwidth]{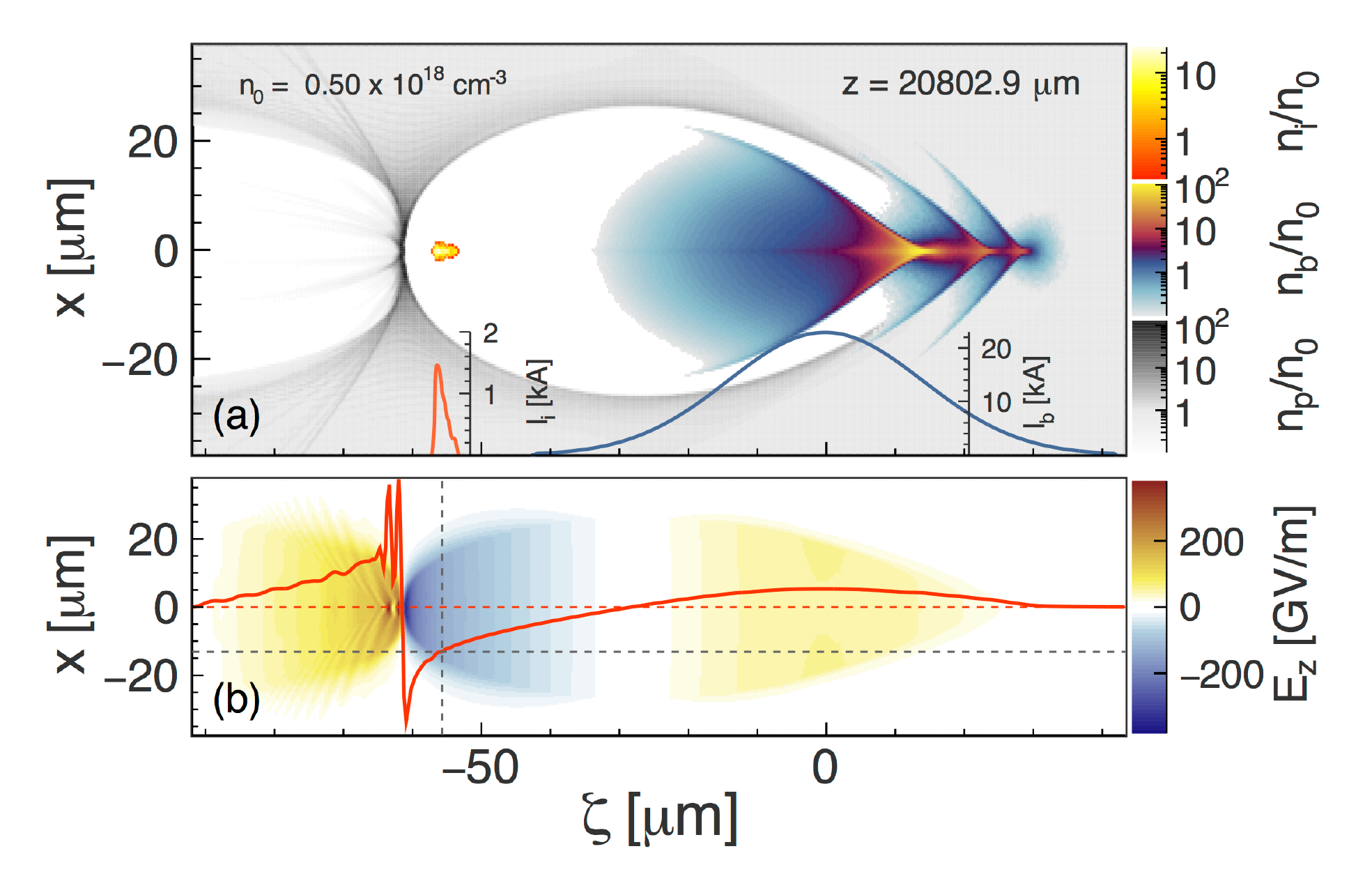}
  \caption{
    Same simulation as in Figure~\ref{fig:FACETrake}, but after a distance of $20.8~\mathrm{mm}$ of propagation in plasma.
    (a) Spatial particle density. 
    Here a short bunch of $8.77~\mathrm{pC}$ has been injected and subsequently accelerated up to more than $2.5~\mathrm{GeV}$ energy.
    (b) Longitudinal electric fields. The bunch experiences an accelerating gradient largely beyond $100~\mathrm{GV/m}$.   
  }
\label{fig:FACETrakeprop} 
\end{figure}

The next example considers electron bunches 
similar to those provided by FACET at SLAC. 
These beams~\cite{Hogan:FACET} can be approximated by Gaussian longitudinal ($\sigma_z = 14~\mathrm{\mu m}$) 
and transverse ($\sigma_x = \sigma_y = 10~\mathrm{\mu m}$) profiles 
with peak currents of $23~\mathrm{kA}$, transverse normalized emittances of 
$\epsilon_{n,x} = 50~\mathrm{\mu m}$ and $\epsilon_{n,y} = 5~\mathrm{\mu m}$,
and an energy of $23~\mathrm{GeV}$ with a relative spread of $1\%$. 
The most important difference with respect to the FLASHForward case in Sec.~\ref{sec:FLASHForward} is the 
significantly higher peak current of the driver beam $\Lambda_b^0 = 2.7$.
Figure~\ref{fig:FACETrake} shows a simulation of the FACET beam transversing a 
$n_0 = 5\times10^{17}~\mathrm{cm}^3$ density plasma doped with He at $n_{\mathrm{He}} = 0.002~n_0$,
within a gas jet of $L_{\mathrm{He}}=100~\mathrm{\mu m}$ diameter.
The fact that the blowout is bigger in this case, 
implies that the maximum difference of the wake potential $\Delta\psi_{\mathrm{max}}$ 
is bigger in the focusing region of the ion cavity (cf. Figure~\ref{fig:boutscal}).
This allows for more flexibility for the selection of the HIT species, 
provided that the difference between $E_z^{\mathrm{I}}$ and $E_z^b$ is equivalently greater.
This also means that the HIT species can be chosen in a way that ionization happens closer to the end of the cavity,
where the blowout radius gets smaller so that the injection volume is more uniform and constrained.
Figure~\ref{fig:FACETrake}(b) shows the ionization volume within the FACET beam blowout, on top of the wake potential contours. 
Here, the ionization volume is basically a part of the trapping region, 
but, as discussed before, the electrons with large initial offsets with respect to the axis escape the blowout region.
A sufficient condition for trapping of an electron is that it
reaches a $\psi_f$ contour before escaping the blowout in straight backward propagation. 
The maximum radius fulfilling this specific condition is
$r_{\mathrm{max}}\approx 1.6~k_p^{-1} = 12~\mathrm{\mu m}$ (cf.~Figure~\ref{fig:FACETrake}(b)).
This allows the volume of injection to be estimated  
$V_{\mathrm{inj}} \simeq \pi r_{\mathrm{max}}^2~\Delta\zeta_\mathrm{ion}$ and hence of the total trapped charge 
$Q_{\mathrm{He}} \simeq -e n_{\mathrm{He}}~\pi r_{\mathrm{max}}^2~L_{\mathrm{He}} = 7.2~\mathrm{pC}$.
Figure~\ref{fig:FACETrakeprop} shows a short bunch of $0.8~\mathrm{\mu m}$~(rms) length 
and $8.77~\mathrm{pC}$ charge, propagating with an accelerating field of $E_z \approx 130~\mathrm{GV/m}$ (Figure~\ref{fig:FACETrakeprop}(b)).
At this point the injected bunch has been propagating for $20.8~\mathrm{mm}$ in the plasma wake.
\begin{figure}[!t]
 \centering
  \includegraphics[width=1.0\columnwidth]{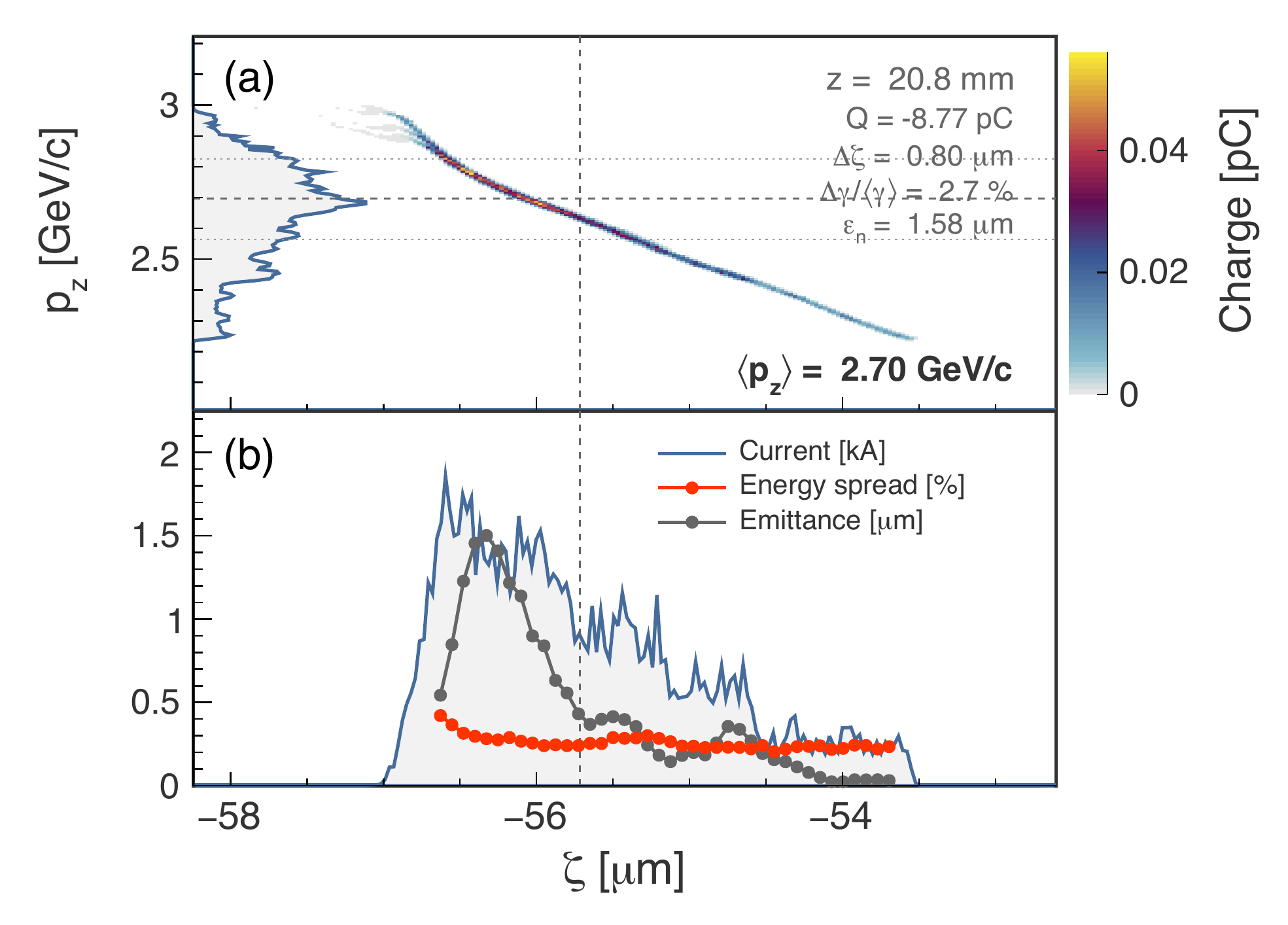}
  \caption{
    Witness bunch properties after $20.8~\mathrm{mm}$ of acceleration.
    (a) shows the absolute charge distribution of the bunch in longitudinal phase space ($p_{z}$ vs. $\zeta$ plane).
    The projection of this distribution in $p_{z}$ is depicted on the left axis.
    (b) displays the bunch-current dependence on $\zeta$.
    The relative energy spread and the transverse emittance 
    are plotted for different longitudinal slices along the bunch.   
  }
\label{fig:FACETrakebunch} 
\end{figure}
The bunch properties are shown in Figure~\ref{fig:FACETrakebunch} in more detail.
The longitudinal phase space (Figure~\ref{fig:FACETrakebunch}(a)) exhibits a linear chirp with an average energy of $\sim2.6~\mathrm{GeV}$, a total relative energy spread of~$6\%$
and a relative energy spread in FWHM of~$2.7\%$. 
The sliced bunch properties can be seen in more detail in Figure~\ref{fig:FACETrakebunch}(b).
The current profile has a maximum at the tail of the bunch of $\sim1.5~\mathrm{kA}$ 
and linearly decays towards its front~(Figure~\ref{fig:FACETrakebunch}(b)).
The relative energy spread ($\sim0.3\%$), and the normalized transverse emittance ($\le 1.5~\mathrm{\mu m}$)
are shown for different slices in $\zeta$.
Considering $r_{\mathrm{max}}= 12~\mathrm{\mu m}$ and Eq.~(\ref{eq:rakeemit}), the estimated normalized emittance is $1.4~\mathrm{\mu m}$,
in excellent agreement with the simulation.
Further acceleration of the captured beam is possible until the driver is energy-depleted. 
A simple estimate yields the maximum achievable bunch energy when considering 
the decelerating gradients experienced by the driver beam of $50~\mathrm{GV/m}$~(cf. Figure~\ref{fig:FACETrakeprop}(b)). 
This limits the driver propagation distance to $\sim 46~\mathrm{cm}$, and hence, 
the maximum energy of the injected beam to about $46~\mathrm{GeV}$ assuming an acceleration of the trailing bunch 
at a continuing rate of $100~\mathrm{GV/m}$. 

\section{Beam loading}
\label{sec:beamload}
Ionization-based injection methods have the ability of tuning the amount of injected charge
by means of the dopant species concentration and/or the length of the injection section 
(approximately the gas-jet diameter in the here considered case). 
In the two previous examples the charge of the injected beam was chosen to be low, so that the beam
barely deforms the wakefields, and the energy chirp basically is dominated by the local shape of the unloaded wakefield $E_z$.
In general, an electron beam injected in the accelerating region of the blowout cavity
deforms the slope of the longitudinal field with respect to the unloaded case.
For a given witness-beam current profile $\Lambda_w$, this effect can be calculated from Eq.~(\ref{eq:boeqE}).
In this discussion, we assume that the bunch is trapped close to, but before the end of the cavity. 

\begin{figure}[!t]
 \centering
  \includegraphics[width=1.0\columnwidth]{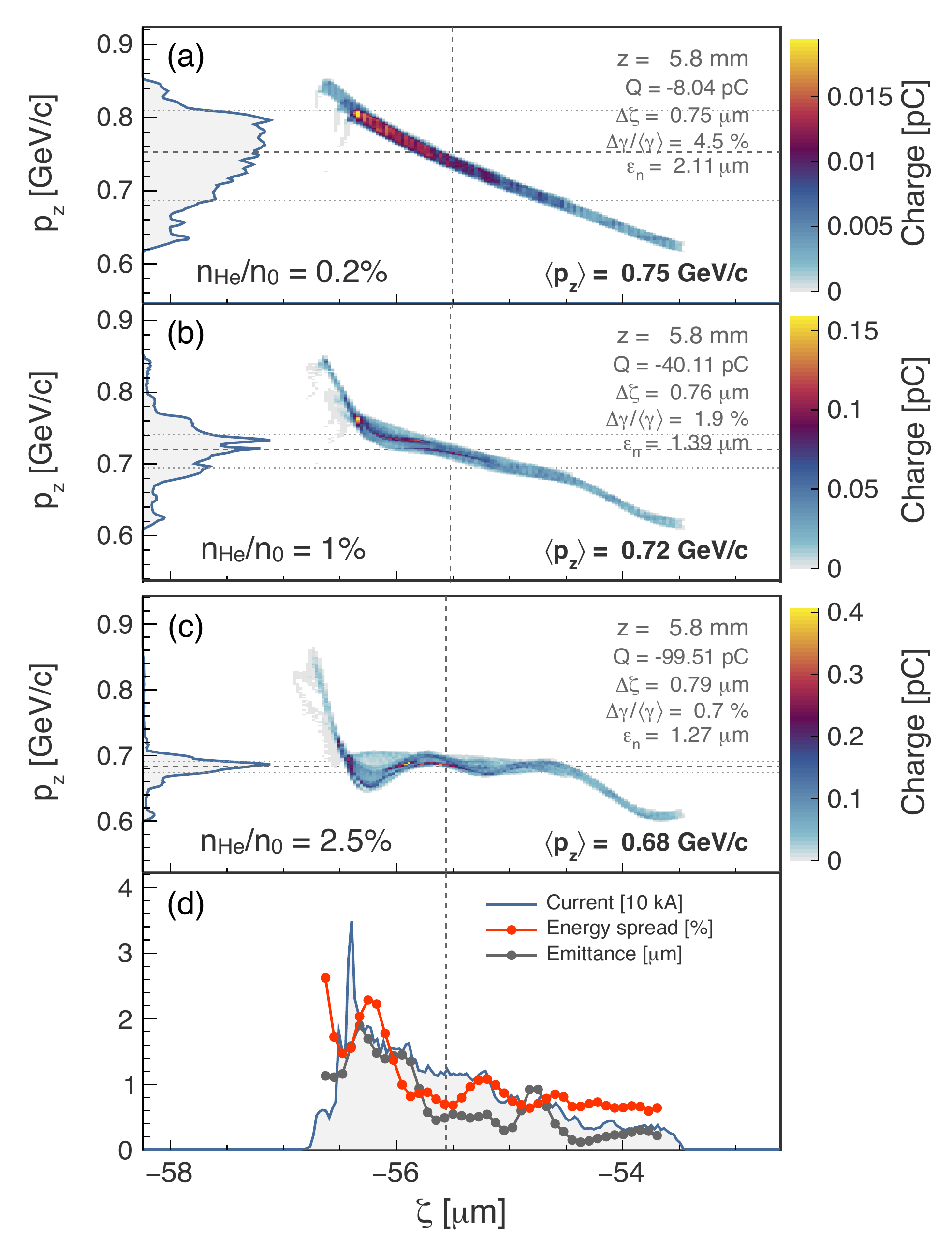}
  \caption {
    Longitudinal phasespace of the witness bunches after acceleration over a distance of $5.8~\mathrm{mm}$,
    for the unloaded case (a), a partially loaded one (b) and the optimized beam loading case (c).
    The projection of these distribution in $p_{z}$ are depicted on the left axis.
    The bunch-current profile, the relative energy spread and the transverse emittance 
    are plotted in panel (d) for various longitudinal slices along the optimized bunch.   
  }
\label{fig:FACETrakebunchbl} 
\end{figure}

As discussed in Sec.~\ref{sec:wiii} for high-current driver beams, 
the velocity of the electrons in the plasma sheath approaches the speed of light in the backwards direction ($\beta_{lz} \approx -c$)
when electrons in the sheath reach the point of maximal blowout radius.
From this point on, the plasma electrons in the sheath are accelerated in the forward direction and
the velocity (third) term in Eq.~(\ref{eq:boeqE}) only dominates at the very end of the bubble,
when the electrons velocity approaches the speed of light in the right direction ($\beta_{lz} \approx c$).
In the following discussion, the sheath velocity term in Eq.~(\ref{eq:boeqE}) was implicitly neglected,
considering that the driver bunch is situated at a point where the electrons in the plasma sheath
do not have a sufficiently high positive longitudinal velocity to contribute significantly.
\begin{figure}[!b]
 \centering
  \includegraphics[width=1.0\columnwidth]{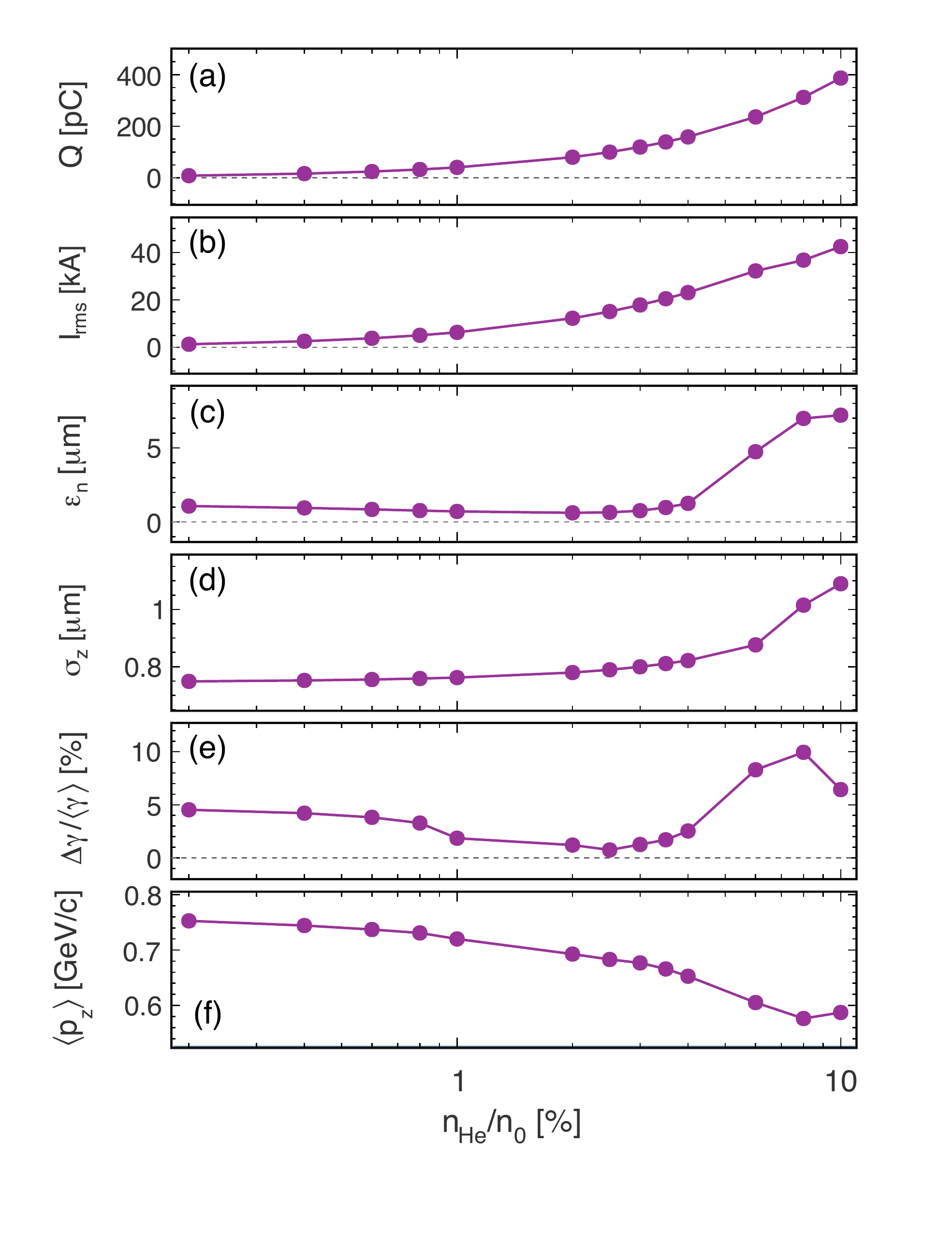}
  \caption{
    Injected bunch properties after a distance of $5.8~\mathrm{mm}$ of acceleration 
    as a function of the dopant He concentration.
    (a) Total charge. (b) Current (rms). (c) Normalized emittance. (d) Length (rms). 
    (e) Relative energy spread (FWHM). (f) Average longitudinal momentum.  
  }
\label{fig:FACETbeamload} 
\end{figure}
With this approximation Eq.~(\ref{eq:boeqE}) reads
\begin{equation}
\frac{\partial_\zeta E_z}{k_p E_0} =  - \frac{2\Lambda_w}{(k_p r_l)^2} + \frac{4}{(k_p r_l)^2} \left[\frac{E_z}{E_0}\right]^2 \,. \label{eq:boeqEbl}
\end{equation}
This equation allows the current needed to flatten the slope of $E_z$
along the beam to be estimated.
Assuming perfect flattening, e.g. the left hand side of Eq.~(\ref{eq:boeqEbl}) to be identically zero gives
\begin{equation}
\Lambda_w = 2 \left[\frac{E_z^w}{E_0}\right]^2 \,, \label{eq:beamloading}
\end{equation}
where $\Lambda_w$ refers to the normalized charge per unit length of the witness bunch,
and $E_z^w$ is $E_z$ at the position of the witness bunch $\zeta_w$.
This is a local criterion on the slope of $E_z$. The shape of $E_z$ along the bunch is obtained 
through integration of Eq.~(\ref{eq:boeqEbl}).
This has been done by M. Tzoufras et al.~\cite{Tzoufras:2008vr} 
for certain bunch-current profiles that permit for an analytical solution. 
Of particular interest are bunch profiles that totally flatten the field slope along the bunch.
These are found to be trapezoidal profiles with the maximum located at the leading edge of the beam at $\zeta_w$,
linearly decaying towards the tail
\begin{equation}
\Lambda_w = \sqrt{\left[\frac{E_z^w}{E_0}\right]^4+\left[\frac{k_pr_m}{2}\right]^4} - \frac{E_z^w}{E_0}~k_p(\zeta-\zeta_w)\,. \label{eq:bltraptzou}
\end{equation}
Generally, the witness bunches generated by WII injection do not have this ideal trapezoidal shape for an optimally tailored beam loading. 
In the following we explored how the energy chirp of the witness bunch changes in respect to the unload case in Fig.~\ref{fig:FACETrakebunch}, when the injected charge is increased.
Figure~\ref{fig:FACETbeamload}, presents the parameters of injected bunches 
for a series of simulations identical other than the concentration of the dopant He in the gas-jet.
The simulation results show witness beam parameters at a distance of $5.8~\mathrm{mm}$ downstream of the location of the injection.
This is substantially before energy depletion of the driving beam occurs, and the energy gain of the witness beam is therefore limited to about $600-750~\mathrm{MeV}$.
The data for the lowest He density, $n_{\mathrm{He}}=0.002~n_0=10^{15}~\mathrm{cm}^{-3}$,
in Figure~\ref{fig:FACETbeamload} corresponds to the bunch obtained in the simulation presented in Sec.~\ref{sec:FACET}. 
As the concentration of He is increased, both the total charge and the rms current of the bunch 
increase accordingly (see Figure~\ref{fig:FACETbeamload}(a) and (b)). 
In contrast, the normalized emittance $\epsilon_n$ (Figure~\ref{fig:FACETbeamload}(c)) 
and the rms bunch length $\sigma_z$~(Figure~\ref{fig:FACETbeamload}(d)) are not changed significantly for increasing bunch charge. 
This results from both the normalized emittance and the rms length of the injected bunch depending primarily on the volume of injection,
so that increasing the He concentration does not have a severe impact.
On the contrary, the projected energy spread of the bunch in full width at FWHM~(Figure~\ref{fig:FACETrakebunch}) 
is dominated by the slope of $E_z$ along the bunch, which in turn 
depends on the current profile of the injected beam~(Eq.~(\ref{eq:boeqEbl})).
Therefore, a clear reduction of the projected FWHM energy spread can be observed (see Figure~\ref{fig:FACETbeamload}(e)) for increasing charge of the injected bunch. 
The energy spread is minimal for a dopant density of about $n_{\mathrm{He}} = 0.025~n_0$ (cf. Figure~\ref{fig:FACETbeamload}(e)). 
For this case, the injected bunch has a total charge of $~100~\mathrm{pC}$, 
and a rms current of $I_{\mathrm{rms}}\approx15~\mathrm{kA}$ (Figure~\ref{fig:FACETrakebunchbl}(d)). 
Both the bunch length $\sigma_z = 0.8~\mathrm{\mu m}$ and the normalized emittance $\epsilon_n = 1.3~\mathrm{\mu m}$
remain at the same magnitude as in the unloaded case. 
If further increased, the charge of the bunch starts to overload and severely deform the $E_z$ slope.
Figure~\ref{fig:FACETrakebunchbl} shows the energy chirp for the unloaded case with $n_{\mathrm{He}}/n_0 = 0.2~\%$ (Figure~\ref{fig:FACETrakebunchbl}(a)),
a partially loaded one with $n_{\mathrm{He}}/n_0 = 1.0~\%$ (Figure~\ref{fig:FACETrakebunchbl}(b)) and the best loaded case with $n_{\mathrm{He}}/n_0 = 2.5~\%$~(Figure~\ref{fig:FACETrakebunchbl}(c)).
More details on the current profile, the sliced energy spread and normalized emittance can be seen in Figure~\ref{fig:FACETrakebunchbl}(d),
for the optimum beam loading case. 
The peaked projected energy spectrum in Figure~\ref{fig:FACETrakebunchbl}(c)
has a relative energy spread (in FWHM) of only $0.7~\%$, while in the unloaded case of Figure~\ref{fig:FACETrakebunchbl}(a),
the same quantity amounts up to $4.5~\%$.
This means a reduction of $\sim85\%$ for optimal beam loading in respect to the unloaded case. 
The longitudinal phase space correlation is small for the high-current part of the beam, 
which is accelerated approximately at a constant rate along the intra-bunch axis $E_z^w \approx 117~\mathrm{GV/m}$.

We compare this result with the theoretical estimations given in Eq.~(\ref{eq:beamloading}) and Eq.~(\ref{eq:bltraptzou}).
Using $E_z^w \approx 117~\mathrm{GV/m}$ in Eq.~(\ref{eq:beamloading})    
leads to a beam current of $I_w \approx 50~\mathrm{kA}$, 
while using Eq.~(\ref{eq:bltraptzou}) with $k_pr_m = 3.3$ (Eq.~(\ref{eq:radiuslu})),
one obtains $I_w \approx 34~\mathrm{kA}$. 
The current profile of the bunch in the PIC simulation (Figure~\ref{fig:FACETrakebunchbl}(d)) has peak currents
on the order of $30~\mathrm{kA}$, which is not far from these analytical estimations.
Based on these results, and on the applicability of the scalings for $k_pr_m$ and $E_z^b$
(Eqs.~(\ref{eq:radiuslu}) and~(\ref{eq:ezbmax})) with the peak current of the drive beam,
we find the following expression relating the peak current of the witness and the driver beam
with the actual driver-to-witness energy transformer ratio $R_w$
\begin{equation}
\frac{I_w}{I_b} \approx \sqrt{\left(\frac{R_w}{\sqrt{2}}\right)^4+1}\,.\label{eq:transcurr}
\end{equation}
Since WII injection requires high-current drivers with $I_b \gtrsim 8.5~\mathrm{kA}$ and the witness
bunches are injected at a position yielding a high transformer ratio,
Eq.~(\ref{eq:transcurr}) implies that the produced bunches feature currents of tens of $\mathrm{kA}$
resulting in an optimal beam loading and the consequent reduction of the energy spread.
Since in addition the WII injected bunches are naturally of low emittance (Eq.~(\ref{eq:wiibunchprop})),
the latter result implies that the WII produced bunches have a high normalized brightness
$B_n \propto I_b/\epsilon_n^2$, which is comparable and might be superior to electron beams generated in
state of the art facilities for FEL application (LCLS, XFEL, etc.),
when operating at plasma densities of $n_0=10^{18}~\mathrm{cm}^{-3}$ or higher.

\section{Self-similar Staging}
\label{sec:sss}
As a consequence of this study,
provided that the WII injected witness bunches are short, of low-emittance and of high-current,
they fulfil the essential requirements to drive strong plasma waves in a higher density plasma.
The field configuration in this new plasma stage, in which the witness beam acts as the driver,
may again trigger WII injection from an appropriate dopant species.
Given the driver in the first plasma stage was near to the resonance condition, $k_p\sigma_z \approx 1$, 
and the length of the produced witness is $k_p \sigma_z \sim 0.1$ (Eq.~(\ref{eq:wiibunchprop})), 
the density in the next plasma stage, in which the witness drives a resonant blowout wake, 
needs to be about $100$ times greater than in the first stage.
We can therefore formulate the following relation 
$n_0'/n_0 = (\sigma_z^b/\sigma_z^{w})^2 \approx 100$,
where $n_0$ and $n_0'$ are the densities of the first and second stage, respectively.
Owing to the increased density, the witness bunches generated in the second plasma stage 
(where the previous witness bunch acts now as driver) will have a dramatically reduced size and normalized emittance.
Using the scaling of the witness bunch properties (Eq.~(\ref{eq:wiibunchprop})), 
one can estimate a reduction of one order of magnitude in size and normalized emittance of the second witness beam with respect to the first, 
and an about two orders of magnitude increased normalized brightness ($B_n \propto I_b/\epsilon_n^2$).
In addition, due to the high transformer ratio, 
the energy per electron of the second witness can be double or thrice the one in the first witness.
This means, an increase of up to one order of magnitude in the energy per electron of the second witness with respect to the first driver.     
This is a novel concept of self-similar staging in PWFA,
in which every next stage operates in a significantly increased plasma density.

We have explored the validity of this new concept of staging using PIC simulations in the same manner
as we did in Secs.~\ref{sec:FLASHForward}~and~\ref{sec:FACET}.
In this case we used as driver an electron beam with exactly the same statistical properties as the
witness beam in the beam-loaded FACET case (Figure~\ref{fig:FACETrakebunchbl}~(c) and (d)).
The field structure of this driver operating in a plasma with $n_0 = 4\times10^{19}~\mathrm{cm}^{-3}$ density
allows for WII injection from an appropriate ionization level at the required ionization threshold.
In this particular example, the fifth ionization level of neon,
with an ionization threshold of $E_\mathrm{ion} = 557~\mathrm{GV/m}$ fulfills the condition expressed in Eq.~(\ref{eq:ionth}) (cf.~Figure~\ref{fig:adkprob}).
The witness bunches generated in this second plasma stage feature ultra-short durations ($100~\mathrm{as}$ (rms)),
and ultra-low normalized emittances ($60~\mathrm{nm}$), while exhibiting high-currents ($15~\mathrm{kA}$) for a proper beam loading.
Provided that the normalized emittance of the second witness is about $20$ times smaller than the first witness,
and that the peak current is essentially the same,
the peak normalized brightness of the second witness bunch in respect to the first is about $400$ times higher.
Such values would largely exceed those from state of the art conventional linear electron accelerators.
Furthermore, provided that the beam is injected in a phase position such that the transformer ratio is around $3$,
its final energy after two stages could exceed $100~\mathrm{GeV}$ per electron,
if the self-similar staging process is initiated by the $23~\mathrm{GeV}$ FACET electron beam.

The same concept of staging can be applied to electron bunches produced in laser-driven plasma wakefield accelerators (LWFA),
as they could have enough current~\cite{wiggins:2010,lundh:2011} to drive strong plasma wakes
and trigger WII injection in a second plasma stage of increased density (PWFA).
A similar hybrid staging concept was proposed as a way of producing, from a LWFA stage,
driver/witness electron beam pairs that can operate a PWFA stage
at higher plasma densities than driver/witness pairs from conventional accelerators~\cite{Hidding:2010prl}.
Here, we propose to use just one high-current LWFA produced electron beam 
that drives strong plasma wakes at (or near) the resonant density,
and injects its own high-quality witness bunch by means of the WII injection mechanism. 
The requirements for WII injection are mainly demanding on the current profile of the beam,
but not too strict in terms of emittance and energy spread.
This makes electron beams generated in LWFA experiments suitable candidates
to drive the WII injection mechanism in a significantly increased density plasma stage,
for the creation of ultra-short, low-emittance and high-current electron beams
with multi-GeV energy in a room the size of a laser lab.

\section{Summary}
\label{sec:summary}
The WII injection technique requires high-current ($I_b \gtrsim 8.5~\mathrm{kA}$) 
and moderate length ($L_b \approx \lambda_p$) drive beams,
to generate plasma waves in a strong blowout regime.
In this regime the plasma wake is capable of trapping electrons from ionization.
The drive beams need to be smaller transversely in the high-current region than the blowout radius ($k_p\sigma_r < k_pr_m \approx 2\sqrt{\Lambda_b}$) 
to excite the plasma wave most efficiently and to transport the beam, 
but wide enough to obtain a maximum radial electric field below the ionization threshold
in the region of the driver ($k_p\sigma_r>0.63~\sqrt{\Lambda_b}$).
Moreover, the driver's beta function needs to be greater than the matched beta 
in the focusing ion channel, so that the driver beam is transversely compressed after the injection
to ensure a stable propagation that mitigates the head erosion.
This sets a condition for the beam transverse emittance $k_p\epsilon_n < \sqrt{\gamma/2}~(k_p\sigma_r)^2$.
By a proper choice of the HIT dopant species, 
the wakefield induces injection and trapping from a restricted phase-space area at the back of the first plasma bucket. 
The electron bunches injected by means of the WII injection method have, by construction, 
a characteristic size and a normalized emittance of the order of a fraction of the plasma skindepth.
This means sub-micron length and emittance for densities close to $10^{18}~\mathrm{cm}^{-3}$.
Because the phase of injection is located near to the end of the cavity, 
the WII injected witness beams are accelerated at (or close to) the optimal phase of the plasma wake, 
where the accelerating gradients are the highest and the transverse fields are focusing.
The requirements for WII injection comply with those for an efficient operating mode of the PWFA~\cite{lotov:053105},
providing the best driver-to-witness energy exchange (i.e. the highest transformer ratio).
This injection strategy results in a controlled beam loading to flatten the accelerating field
along the witness bunch length, which leads to a significant reduction of the energy spread.

The required witness currents for optimized beam loading are on the order of tens of $\mathrm{kA}$.
As a result, the WII injection can produce high-quality electron bunches with short pulse length, low normalized emittance, 
high-current, high-brightness and low-energy spread, with an energy per electron of around three times that of the drive beam.
This can be achieved in a relatively simple experimental setup. 

The witness bunches generated by the above process fulfil all requirements to again trigger
WII injection as driver beams in substantially higher-density plasmas.
This new concept of self-similar staging has the potential
to produce electron beams with unprecedented energy and quality in PWFA.

\begin{acknowledgments}
We thank the OSIRIS consortium (IST/UCLA) for access to the OSIRIS code. 
Special thanks for support go to J. Vieira and R. Fonseca. 
Furthermore, we acknowledge the grant of computing time by the J\"{u}lich Supercomputing Centre on JUQUEEN under Project No.~HHH23 and the use of the High-Performance Cluster (IT-HPC) at DESY. 
This work was funded by the Humboldt Professorship of B. Foster, the Helmholtz Virtual Institute VH-VI-503 and ARD program. 
\end{acknowledgments}

\bibliography{WIII}

\end{document}